\documentclass[preprint, preprintnumbers,showpacs, showkeys, aps, prd,
  amsmath, amsfonts, amssymb, eqsecnum, nofootinbib, floatfix,
  secnumarabic]{revtex4-1} 
\usepackage{amsmath,amssymb,graphicx}
\usepackage{slashed}
\usepackage{epsf}
\usepackage{float}
\pdfoutput=1

\newcommand{\nc}{\newcommand}
\nc{\postscript}[2]
{\setlength{\epsfxsize}{#2\hsize}\centerline{\epsfbox{#1}}}
\nc{\non}{\nonumber}
\nc{\hc}{\hbox {h.c.}} \nc{\re}{\hbox {Re}} 
\nc{\mev}{\hbox {MeV}} \nc{\gev}{\;\hbox {GeV}} \nc{\tev}{\;\hbox {TeV}}
\def\lsim{\mathrel{\raise.3ex\hbox{$<$\kern-.75em\lower1ex\hbox{$\sim$}}}}
\def\gsim{\mathrel{\raise.3ex\hbox{$>$\kern-.75em\lower1ex\hbox{$\sim$}}}}

\nc{\etal}{{\it et al.}}
\nc{\Lsp}{\;\;\;\;\;\;\;\;\;\;}  \nc{\LLLsp}{\lspace \lspace}
\nc{\lsp}{\;\;\;\;\;\;}
\nc{\spac}{\;\;\;}
\nc{\noi}{\noindent}
\nc{\beq}{\begin{equation}}   \nc{\eeq}{\end{equation}}
\nc{\bea}{\begin{eqnarray}}   \nc{\eea}{\end{eqnarray}}
\nc{\baa}{\begin{array}}      \nc{\eaa}{\end{array}}
\nc{\bit}{\begin{itemize}}    \nc{\eit}{\end{itemize}}
\nc{\ben}{\begin{enumerate}}  \nc{\een}{\end{enumerate}}
\nc{\bce}{\begin{center}}     \nc{\ece}{\end{center}}

\nc*{\plots}{.}%
\nc*{\diagrams}{.}%

\def\ie{{\it i.e.,}}

\def\sq2{\sqrt{2}}

\def\ph{\varphi}

\def\m4{m^4(\ph)}
\def\mn2{m_n^2}

\def\v5{V^{(5)}}

\def\baa{\begin{array}}
\def\eaa{\end{array}}

\begin{document}

\begin{flushright}
 \mbox{\normalsize \rm CUMQ/HEP 182}\\
\end{flushright}

\vskip -10pt

\title{Flavor-changing decays of the top quark in 5D Warped Models}

\author{Alfonso D\'{i}az-Furlong$^1$\footnote{alfonso.furlong@correo.buap.mx}}
\author{Mariana Frank$^2$\footnote{mariana.frank@concordia.ca}}
\author{Nima Pourtolami$^2$\footnote{n\_pour@live.concordia.ca}}
\author{Manuel Toharia$^{2,3}$\footnote{mtoharia@dawsoncollege.qc.ca}}
\author{Reyna Xoxocotzi$^4$\footnote{xoxo\_reyna@yahoo.com.mx}}
\affiliation{$^1$ Facultad de Psicolog\'{i}a, Benem\'{e}rita Universidad Aut\'{o}noma de Puebla, 4 sur, Centro Hist\'{o}rico, Puebla, Pue., M\'{e}xico, C.P. 72000}
\affiliation{$^2$ Department of Physics, Concordia University,
7141 Sherbrooke St. West, Montreal, Quebec, Canada H4B 1R6}
\affiliation{$^3$ Physics Department, Dawson College,
 3040 Sherbrooke St., Westmount, Quebec, Canada H3Z 1A4}
\affiliation{$^4$ Facultad de Ciencias F\'{i}sico-Matem\'{a}ticas, Benem\'{e}rita Universidad Aut\'{o}noma de Puebla, Apdo. Postal 1364, C.P. 72570, Puebla, Pue., M\'{e}xico} 


\begin{abstract}

We study flavor changing neutral current decays of the top
quark in the context of general warped extra dimensions, where the
five dimensional metric is slightly modified from 5D anti-de-Sitter
(AdS$_5$). These models address the Planck-electroweak hierarchies of
the Standard Model {\it and} 
can obey all the low energy flavor bounds and electroweak precision tests,
while allowing the scale of new physics to be at the TeV level, and
thus within the reach of the LHC at Run II.  We perform the 
calculation of these exotic top decay rates for the case of a bulk
Higgs, and thus include in particular the effect of the additional
Kaluza-Klein (KK) Higgs modes running in the loops, along with the
usual KK fermions and KK gluons. 
\end{abstract}

\pacs{11.10.Kk, 12.15.Ff, 14.60.Pq}

\maketitle

\section{Introduction}
\label{sec:intro}
Extra-dimensional models with warped space-time geometry provide a simple and elegant way
to understand the hierarchy problem. In the original models (Randall
Sundrum, or RS \cite{Randall:1999vf}), the Standard Model (SM) is
embedded in a slice of anti de Sitter space (AdS$_5$) with two
manifolds bounding the extra dimension: one at the  Planck scale, the
other at the TeV scale. Warping induces an exponential hierarchy
between the effective cutoff scales of the theory at the two
manifolds. The  smallness of the electroweak symmetry breaking (EWSB)
scale emerges due to a low cutoff near the TeV brane, while the high
scale of gravity is generated at the other end. If all the SM fields
live on the TeV brane, the model cannot solve the SM flavor puzzle,
and higher dimensional operators induce large flavor-changing neutral
currents (FCNC) disallowed by the low energy data. Allowing the SM
fermion fields to leak into the bulk \cite{bulk} can help resolving the flavor hierarchy
problem, but is not always sufficient to protect RS from severe flavor
and electroweak constraints \cite{constraints}.
The reason is that in these models, the interactions of ordinary quarks with the KK gauge bosons
are non-universal in flavor, which  induces tree level FCNC processes
mediated by these heavy gauge bosons. Constraints from the CP violating observable in the kaon system,
$\epsilon_K$, result in  generic bounds on the mass of the lightest KK gauge
boson excitation (KK gluon) of 10-20 TeV.  Moreover, because of the mixing of
the KK gauge bosons with the SM $Z$ boson within EWSB, the $Z$
couplings to quarks become flavor non-universal, producing dangerous 
contributions to electroweak precision observables. With such heavy KK
masses, there is hardly any hope of seeing such models at the LHC at the present run \cite{lhc}.

There are different resolutions available in the literature to deal
with these constraints. One is to enlarge the gauge field to $SU(3)_c \times
SU(2)_L \times SU(2)_R \times U(1)_X \times P_{LR}$, where the
additional symmetries of $SU(2)_R$ and $P_{LR}$ to the $T$ parameter
offer custodial protection to $Z q_L^i {\bar q}_L^j$ vertices and
lower the constraints on the KK scales to 2-3 TeV
\cite{Agashe:2003zs}. Another possibility is to include brane kinetic
terms for gauge and fermion fields propagating in the bulk, yielding
first KK mode masses of the order of a few TeV \cite{Carena:2004zn},
with precision bounds under control. Another alternative is to introduce a dilatonic scalar to allow for a
deformation of the space-time metric such that it deviates  from the
AdS$_5$ structure in the infrared region (near the TeV brane), while it approaches AdS$_5$
asymptotically in the UV brane 
\cite{Falkowski:2008fz,Batell:2008me,Cabrer:2011fb,Carmona:2011ib}. In the particular model studied
in \cite{Cabrer:2011fb}, the IR brane  is close to a naked metric singularity, outside the
physical interval.  The proximity of the singularity provides a strong
wave-function renormalization for the Higgs field, which suppresses additional contributions to 
the $T$ and $S$ parameters, and can render the theory valid for KK masses
$M_{KK} \sim$ 1-3 TeV. 


With Run II at the LHC, we are entering the precision era for the
SM Higgs physics, but the LHC is also known for being an
effective top quark factory, with millions of top quarks being
produced yearly, and cross sections for pair productions reaching 1
$nb$. The experimental achievements have induced a concerted effort to
improve the theoretical estimates.  Deviations from SM, either in
direct particle production or indirectly through higher order effects,
and/or observables suppressed in the SM constitute areas of useful
examination. Loop-induced dipole operators in
warped space models exhibit a non-trivial dependence on the Higgs
profile, such that the contribution is saturated  as the Higgs approaches the
IR brane, and decreases when the Higgs field is leaking out towards
the UV brane \cite{Delaunay:2012cz}.  Recently, it has been shown
that by including KK excitations of the SM Higgs boson in loop
diagrams (in particular, in those yielding dipole operators of SM
fermions), the effect of summing over enough  KK modes in the brane
limit can add up and increment the value of the amplitudes by some order 1 factors
\cite{Agashe:2014jca}.

FCNC processes of the top quark are extremely suppressed in the SM,
and  in supersymmetry an enhancement is expected in $b \to s \gamma$
rather than $t \to c \gamma$ \cite{Cao:2007dk}, due to allowed values
of $\tan \beta$. In the SM, ${\cal B}(t \to q X) \simeq 10^{-17} -
10^{-12}$, for  $q = c, u$ and $X = Z,  g, \gamma, H$
\cite{AguilarSaavedra:2004wm}\footnote{Specifically,  ${\cal B}(t \to
  c \gamma)=\left (4.6^{+1.2}_{-1.0} \pm 0.2 \pm 0.4 ^{+1.6}_{-0.5}
  \right ) \times 10^{-14}$ and ${\cal B}(t \to c g)=\left
  (4.6^{+1.1}_{-0.9} \pm 0.2 \pm 0.4 ^{+2.1}_{-0.7} \right ) \times
  10^{-12}$, while ${\cal B}(t \to c Z)\simeq 1 \times 10^{-14}$ and
  ${\cal B}(t \to c H)\simeq 3 \times 10^{-15}$ \cite{AguilarSaavedra:2004wm}.}. Thus these decays
are suppressed in the SM and  indicate that any significant
enhancements  could signal New Physics effects.   

Models have been designed where ${\cal B}(t \to c \gamma)$ can reach
$10^{-12}-10^{-7}$. In models with extra quarks, ${\cal B}(t \to q
Z)\simeq 1.1 \times 10^{-4}$, ${\cal B}(t \to q H)\simeq 4.1 \times
10^{-5}$, and ${\cal B}(t \to q \gamma)\simeq 7.5 \times 10^{-9}$,
${\cal B}(t \to q g)\simeq 1.5 \times 10^{-7}$ \cite{AguilarSaavedra:2002kr}.
In Two-Higgs Doublet Models which violate FCNC at tree level, ${\cal B}(t \to c
H)\simeq 1.5 \times 10^{-3}$, ${\cal B}(t \to u H)\simeq 5.5 \times
10^{-6}$, and the radiative decays ${\cal B}(t \to c Z)\simeq
10^{-7}$, ${\cal B}(t \to c \gamma) \simeq 10^{-6}$ and ${\cal B}(t
\to c g)\simeq  10^{-4}$  \cite{Atwood:1996vj}.
In Two-Higgs doublet models ${\cal B}(t \to
c Z)\simeq  \times 10^{-10}$, ${\cal B}(t \to c \gamma)\simeq  \times
10^{-9}$,  ${\cal B}(t \to c g)\simeq   10^{-8}$ and ${\cal B}(t \to c
H) \simeq 10^{-5}$ \cite{Luke:1993cy}. In MSSM, the largest results
are obtained assuming  non-universal squark masses, and these are
${\cal B}(t \to q Z)\simeq 2 \times  10^{-6}$, ${\cal B}(t \to q
\gamma)\simeq 2 \times 10^{-6}$,  ${\cal B}(t \to q g)\simeq  10^{-4}$
and ${\cal B}(t \to q H) \simeq 10^{-5}$ \cite{Cao:2007dk}. 
The decay $t \to c \gamma$ was also evaluated in
 top color assisted technicolor model \cite{Cao:2007bx}, little Higgs models
\cite{Han:2009zm}, models with universal extra dimensions \cite{GonzalezSprinberg:2007zz} and within the context of effective theories \cite{Aranda:2009cd}.  The dipole operators have been explored before in
warped extra dimensions \cite{Csaki:2010aj}, and the decay $t \to c
\gamma$ was investigated in  the context of warped extra dimensional
models in \cite{Gao:2013fxa}, for brane-localized Higgs in models with custodial symmetry. 

The best experimental limits come from searches of FCNC decays at the
LHC. ATLAS \cite{Aad:2015uza}  has published a compilation of limits
based on the full 8 TeV data set at 20.3 $fb^{-1}$. The bounds are:
${\cal B}(t \to q Z) <  \times 0.07(0.08)$ \%, ${\cal B}(t \to u
\gamma)< 0.0161(0.0279)$ \%,  ${\cal B}(t \to c \gamma)< 0.182(0.261)$
\% for the observed(expected) limits. In addition, results from both
ATLAS \cite{Aad:2015uza} and CMS \cite{CMS003}  perform the search in single top
production for the decay $t \to c g$. The most stringent results come
from ATLAS,  yielding ${\cal B}(t \to q g)< 4.0 \times 10^{-5}$ and
${\cal B}(t \to c \gamma) <1.7 \times 10^{-4}$ \cite{Aad:2015uza}. A  review of current experimental constraints and theoretical predictions is presented in \cite{Agashe:2013hma}.

In this work we investigate the contribution to the FCNC top quark decay in a
general warped scenario, which allows a slight modification of the
warping factor along the extra dimension, allowing it to deviate
slightly from the AdS$_5$ metric \cite{Cabrer:2011fb}. This deviation
is such that the warping is more drastic near the TeV brane, while the
background becomes more AdS$_5$-like near the Planck brane.  These
models suppress additional contributions to electroweak precision
variables in the same parameter space region where contributions to
Higgs production cross section \cite{Frank:2013qma} and decay rates
\cite{Frank:2015zwd} are consistent with experimental bounds, and this
is achievable only for bulk Higgs.  We perform the calculation
including fermion profiles consistent with the SM masses and the
CKM quark mixing matrix, and sum over all the fermion and Higgs boson
KK modes in the loops up to the third KK states.
We are particularly interested in the role of the KK excitations of
the Higgs boson.

Our work is organized as follows. In Sec. \ref{sec:model} we introduce
briefly the general warped space model, with emphasis on the Higgs and
fermion zero-mode and KK states. In Sec. \ref{subsec:tree} we analyze
the tree-level decays, while in  Sec. \ref{subsec:analytic} present the
results for the FCNC dipole decay of the top quark. We conclude in
Sec. \ref{sec:conclusion} and leave some of our analytic expressions
for the Appendix (\ref{sec:appendix}).

\section{Warped Space Models with fields in the bulk}
\label{sec:model}
Consider the SM fields propagating in a 5D space with an arbitrary metric
$A(y)$ such that the metric is: 
$$ds^2 = e^{-2A(y)}\eta_{\mu \nu} dx_\mu dx_\nu +dy^2$$ 
where $\eta_{\mu \nu}= diag( -1, 1, 1, 1)$. This  is the most general
ansatz consistent with Minkowski spacetime in 4D. A naked singularity
is located at $y=y_s$ such that the IR brane is located a short
distance from that singularity, at $y=y_1=y_s-\Delta$,  by means of a
stabilizing dilatonic field:  
\begin{equation}
\Phi(y)=- \frac{\sqrt{6}} {\nu} \log \left[\nu^2 k(y_s-y)\right],
\end{equation}
with $k$ the inverse curvature radius of the AdS$_5$ space-time and
$\nu$ a real parameter, corresponding to the metric:  
\begin{equation}
A(y)=ky-\frac{1}{\nu^2}\log \left( 1-\frac{y}{y_s}\right ). 
\end{equation}
The modified AdS$_5$ metric mimics that of the RS models $A(y)=ky$,
for $y \to 0$, while drastically departing from it for $y \to y_s$. In
this model, the hierarchy problem is solved by assuming a Higgs
potential of the form 
\begin{equation}
V(H)=k^2 \left [ a(a-4)-4ae^{\nu \Phi/\sqrt{6}}\right] |H|^2
\end{equation}
We define the CP-even Higgs field as
\begin{eqnarray}
  H(x,y)=\frac{1}{\sqrt{2}} 
  \left (\begin{array}{c} 0\\h(y)+\xi(x,y)\end{array} \right )
\end{eqnarray}
where
 $h(y)$ is the Higgs
background vacuum expectation value (VEV) profile determined by the equations of motion and boundary
conditions given by 
\begin{equation}
h(y)
=h_0 e^{aky} \bigg[1 + (M_0/k -a) \left[ F(y) - F(0) \right]\bigg]\, , 
\end{equation}
where $h_0$ is a normalization factor and $M_0$ is the brane Higgs
mass term (the coefficient of the Higgs boundary potential
$|H|^2\delta(y-y_1)$ at the IR brane) introduced to give rise to the
Higgs zero mode with the correct physical mass. The function $F(y)$
given by 
\begin{equation}\label{Fy}
F(y) =  e^{ - 2 (a-2) k y_s }k y_s \left[ -2(a-2) k y_s \right]^{-1 + 4/\nu^2} \Gamma \left[ 1 - \frac{4}{\nu^2} , -2(a-2) k( y_s - y) \right], 
\end{equation}
is a generalization of the corresponding RS function $F(y) =  e^{ - 2
  (a-2) k y }$. Here $a$ is the parameter that determines the
localization of the Higgs field. The large $a$ limit ($a\gg 2$)
corresponds to a brane-localized Higgs, while for $a$ of order $1$ we
 say that the Higgs is a bulk Higgs field. Note that there is a minimum
value of $a$ that ensures that no new fine-tuning is introduced in the
model in order to solve the hierarchy problem; $a_{min} = 
2$ for RS models, $a_{min} > 2$ for modified AdS$_5$ models
\cite{Cabrer:2011fb,Frank:2013un,Frank:2013qma,Frank:2015zwd}. 

In general, the  oblique precision electroweak parameter $T$ is
enhanced by the compactification 
volume $ky_1$ and it is the most constraining  of the oblique
parameters, while $S$ does not depend on the volume.  
In RS, compatibility with electroweak precision data imposes a lower
bound of around $M_{KK} \gsim 13$ TeV at the 95\% CL \cite{constraints}, bound
which can improve when  the Higgs is delocalized from the IR brane
\cite{Agashe:2008uz}. For $a \ge 2$,  the $M_{KK}$ scale bound becomes $M_{KK} \gsim 7$ TeV at
the 95\% CL. In modified AdS$_5$ models,  the different behavior of the
Higgs profile at the IR brane location $y_1$ results in much more
relaxed bounds on the KK scale. As KK modes are localized towards the
IR brane, their overlapping integrals with the Higgs (and therefore
their contribution to the electroweak parameters $T$ 
and $S$) depend on the values of the physical Higgs wave functions at
the IR brane. The scale of new physics $M_{KK}$ could be as low
as 0.8 TeV \cite{Cabrer:2011fb}, for the Higgs and metric parameters
$a=3.1$ and $\nu=0.5$, while for $\nu\gsim 5$ one starts to recover the RS results.

These models have also been tested by comparing their predictions for
Higgs boson production for bulk Higgs, in the original RS metric and
within a modified metric background \cite{Frank:2013un,Frank:2013qma,Frank:2015zwd}. 
In 5D scenarios with modified AdS$_5$ metric, the results are consistent with the LHC Higgs
 measurements in the same region of the parameter space where flavor
 and precision electroweak constraints are satisfied.  Thus a safe region of parameter
space (minimum UV sensitivity and safe from non-perturbative couplings) exists, requiring
moderate 5D Yukawa couplings $Y^{5D}\sim 1$, as well as  low Higgs
localization parameter values, $a\sim 2-5$.

For fermions, using values for $Y_{5D}\sim 1$ and localizations
coefficients $c$, for the zero mode profiles  for which full
analytical expressions are available \cite{Frank:2013un}, we construct
the Yukawa coupling matrix and the mass matrix with the following elements
\begin{equation}
\label{YYY100}
(y^0_u)_{ij} = \frac{(Y^{5D}_u)_{ij}}{\sqrt{k}} \int_0^{y_1} dy
e^{-4A(y)} h(y) q^{0,i}_L(y)u^{0,j}_R(y)\, , 
\end{equation}
where $Y_u^{5D}$ are the 5D dimensionless Yukawa couplings, $u$
stands for up and down $SU(2)$ singlet quarks, and $q^{0,i}_L(y)$
represent zero mode SM doublets. One can then construct the  KK
profiles for fermions through  solving the differential equations
numerically for all the 6 flavors of the fermion profiles: 
\begin{equation}
\partial_y\left(e^{(2c-1)A(y)}\partial_y\left(
e^{-(c+2)A(y)}\right)\right)f(y)+e^{(c-1)A(y)} m_n^2 f(y) =0
\end{equation}
imposing Dirichlet boundary conditions for the ``wrong'' chirality fermions. 
The overlap integrals along the extra dimension lead to the effective
4D Yukawa coupling matrix, which in the up sector can be  writen as
\begin{equation}
{\bf Y}_{u} = \left(\begin{array}{ccc} 
(y^{0}_{u})_{3\times3}      &  (0)_{3\times 3N}   & (Y^{qU})_{3\times 3N}\\
 (Y^{Qu})_{3N \times3}      & (0)_{3N\times  3N} & (Y_1)_{3N \times 3N}\\
(0)_{3N \times3}      &     (Y_2)_{3N\times  3N} &  (0)_{3N\times 3N}
\end{array}\right),\label{yumatrix}
\end{equation}
with the down sector Yukawa matrix ${\bf Y}_d$ computed in the same
way. The submatrices are obtained by the overlap integrals 
\bea\label{YYY1KK}
Y^{qU}= \frac{(Y^{5D}_u)_{ij}}{\sqrt{k}} \int_0^{y_1} dy e^{-4A(y)} h(y) q^{0,i}_{L}(y)U^{n,j}_{R}(y) \\
Y^{Qu} = \frac{(Y^{5D}_u)_{ij}}{\sqrt{k}} \int_0^{y_1} dy e^{-4A(y)} h(y) Q^{m,i}_{L}(y)u^{0,j}_{R}(y)\\
Y_1 = \frac{(Y^{5D}_u)_{ij}}{\sqrt{k}} \int_0^{y_1} dy e^{-4A(y)} h(y) Q^{m,i}_{L}(y)U^{n,j}_{R}(y) \\
Y_2 = \frac{(Y^{5D^*}_u)_{ij}}{\sqrt{k}} \int_0^{y_1} dy e^{-4A(y)} h(y) Q^{m,i}_{R}(y)U^{n,j}_{L}(y)\, ,
\eea
where the indices $m$ and $n$ track the KK level and $i,j=1,2,3$ are
5D flavor indices. We have included 3 full KK levels so that the Yukawa
matrices in the gauge basis are $21 \times 21$ dimensional matrices.

The  4D effective $21 \times 21$ fermion mass matrix (constructed in a
similar way) is not diagonal due to electroweak symmetry breaking, and
must be diagonalized in order to go to the physical mass basis.  
Once in that basis, we obtain the physical Yukawa couplings by
appropriately rotating the Yukawa matrix in Eq.~(\ref{yumatrix}). As pointed out before,
the Yukawa couplings in the mass basis can receive important corrections in these scenarios
 \cite{Azatov:2009na,Frank:2015sua,Frank:2013un}. Quite
plausibly, these effects could add-up and maybe enhance further
loop-dominated flavor violating (FV) decays of the
top quark, where, similar to Higgs production and decay processes, all KK
excitations (for Higgs and fermions) will contribute.  We investigate this in the following section.

\section{Phenomenology of FCNC decays of the top quark}
\label{sec:pheno}
For the phenomenology section of this work we have considered the
parameter region of the modified $AdS_5$ model with $\nu \simeq 0.5$ 
and  $kL(y_1)\simeq 0.2$, which is the curvature of space at the
location of the IR brane given by 
\bea\label{curvature}
kL(y_1) = { k \Delta_1 \nu^2\over\sqrt{1-2\nu^2/5 + 2k\Delta_1\nu^2+(k\Delta_1)^2\nu^4}},
\eea
with $\Delta_1$ being the distance between the position of the
curvature singularity and the IR brane, $\Delta_1 \equiv y_s -
y_1$. This region of parameter space allows for the lowest possible KK scales of about
$700$ GeV, which are still consistent with the electroweak precision
test parameter bounds \cite{Cabrer:2011fb}. We have considered scenarios with 5 different KK
gluon mass scales at about $\sim 700$ GeV, $1000$ GeV, $1300$ GeV,
$1700$ GeV and $ 2300$ GeV. (Note that constraints from flavor processes
might still force the KK scale to be 1-2 TeV \cite{Cabrer:2011fb}). These masses are achieved through slightly changing the length of the extra
dimension by fixing $A(y_1)$ around  $\simeq 36 - 37$. Once the KK
scale has been fixed, we calculate the 
minimum $a$-parameter (corresponding to the maximally delocalized
Higgs field along the 5th dimension) that satisfies the
constraint  $|F(y_1)| \equiv \delta = 1$, where $F(y)$ is given by
Eq. (\ref{Fy}). This constraint ensures that the Higgs profile solution
which leaks towards the UV brane is still IR dominated, without requiring a fine
tuning of parameters ($M_0/k - a =0$). We have considered 4 values of the
$a$-parameter for each of the KK mass scales mentioned above, all
corresponding to a heavily delocalized bulk Higgs field with $a \in
\{a_{min}, a_{min}+0.5, a_{min}+1, a_{min}+1.5\}$, very
close to the values $a\simeq 3$, $3.5$, $4$ and $4.5$.

Having calculated the lowest $a$-parameter of each model, we scan the
parameter space of the 5D-fermion $c$-parameters and the 5D Yukawa couplings, $Y^{5D}$. The
$c$-parameters correspond to the bulk mass parameter of fermions
(localization of fermions along the 5th dimension) and we need to find
a set of these parameters, $\{c_{q_i}, c_{u_i}, c_{d_i}\}$, for all of
the quark sector fermions ($c_q$ corresponds to the $SU(2)$ doublets,
$c_u$ corresponds to the up-like and $c_d$ to the down-like singlets)
that, combined with our choice of $Y^{5D}$, yield the correct SM quark
masses and CKM mixing angles. For our scan, we consider two orders for
the $5D$ Yukawa couplings, $Y^{5D}\simeq 1$ and $Y^{5D} \simeq
3.$\footnote{For the case of $Y^{5D}\simeq 1$ we still set
 the entry  $(Y^{5D})_{33} \simeq 3$ to be able to achieve the top quark
  mass.} Our approach is such that we randomly choose all the Yukawa
couplings and allow for random deviations from guided ranges for the
$c$-parameters which produce SM-like masses and mixings. (For example,
for UV localized fermions we only scan the range between  $0.55 - 0.7$
and disregard possible points outside of this region.) 
This way we conduct a first estimation of the masses and CKM parameters by
only considering the zero mode fermions, calculating the matrix
elements given in Eq. (\ref{YYY100})   
and filter the results to include points that reproduce values close
to the SM. 

Having fixed the parameters $\nu$, $y_1$, $c$'s, $Y^{5D}$'s and $a$, we
include the KK modes into our previous naive calculation. As mentioned
earlier, we have only considered the full first 3 KK
levels. For a highly delocalized Higgs field considered here,  heavier modes
should decouple fast enough so that the results of considering only
the first 3 KK modes are in good agreement with those of including the
full tower \cite{Frank:2013qma}.
We  solve the differential equations of motion along the 5th dimension to find the
masses and profiles of the zero modes and of the KK modes for all the
quark sector fermions, the gauge bosons, and the Higgs bosons.
Using these profiles we compute the $21\times 21$ Yukawa matrices like
the one shown in Eq. (\ref{yumatrix}) for the up-type quarks.
We  then rotate the quark fields to transform to the mass basis by
diagonalizing the up and down fermion mass matrices given by ${\bf M}_u = {\rm VEV} \times
{\bf Y}_u + M_{KK}$ where $M_{KK}$ is a $21\times 21$ diagonal matrix
whose elements are the masses of the KK modes, $M_{KK} =  diag\{0,0,0,
M^{up}_{Q_1}, \dots, M^{top}_{Q_3}, M^{up}_{U_1}, \dots,
M^{top}_{U_3}\}$ (and similarly in the down sector).

In the physical mass basis, ${\bf M}_u \rightarrow V^u_L {\bf M}_u V^u_R$,  the
Yukawa matrix elements  given by  $ {\bf Y}_u \rightarrow V^u_L {\bf
  Y}_u V^u_R$ are not diagonal, leading to tree-level Higgs mediated FCNCs. To
calculate the CKM matrix, we need to perform the same calculations for
the down sector as well. The CKM matrix is given by $(V_{CKM})_{ij} =
\left(V^u_L V_L^{d \dagger}\right)_{ij}$, where 
$i,j=1,2,3$, i.e., it is the $ 3\times 3$ upper-left corner of the
$21\times 21$ charged current mixing matrix $V^u_L V_L^{d 
  \dagger}$.  At this point, we proceed to the final 
scan check and compare the masses and mixings obtained from the
$21\times 21$ matrices  with those of the SM  and discard 
the phenomenologically inconsistent scan points. We generate in this
way 40 different (and viable) parameter space points for each value of
the Higgs parameter $a$ and each KK mass scale.

\subsection{FV decays of top quark at tree level: $t \to q h$ and  $t \to q Z$ }
\label{subsec:tree}
Using the results of the previous section, we can read off
the coupling strengths of the FCNC decays of the top quark at the tree level. For the $t \to c
h$ decay, these are given by the entries $Y^u_{23}$ and $Y^u_{32}$ and by computing
the following branching ratio \cite{Casagrande:2008hr,Azatov:2009na} 
\begin{equation}\label{Brtqh}
{\cal B} (t \rightarrow q h) = {2 (1 - r_Z)^2 (1 + 2 r_Z)\over (1 -
  r_W)^2 (1 + 2 r_W)} \left( |(Y^u_L)_{q3}|^2 +  |(Y^u_L)_{3q}|^2 +
{12 \sqrt{r_q} r_Z\over (1 - r_Z) (1 + 2 r_Z)}
\text{Re}\left[(Y^u_L)^*_{q3}(Y^u_L)_{3q}\right] \right) 
\end{equation} 
where $r_i \equiv (m_i / m_t)^2$,  $q = 2$ for the branching ratio
of top to charm, and $q = 1$ for top to up.   
  
\begin{figure}[t]
	\vspace{-1cm}
\center
\begin{center}
	\includegraphics[height=10.cm, width=17.5cm]{\plots/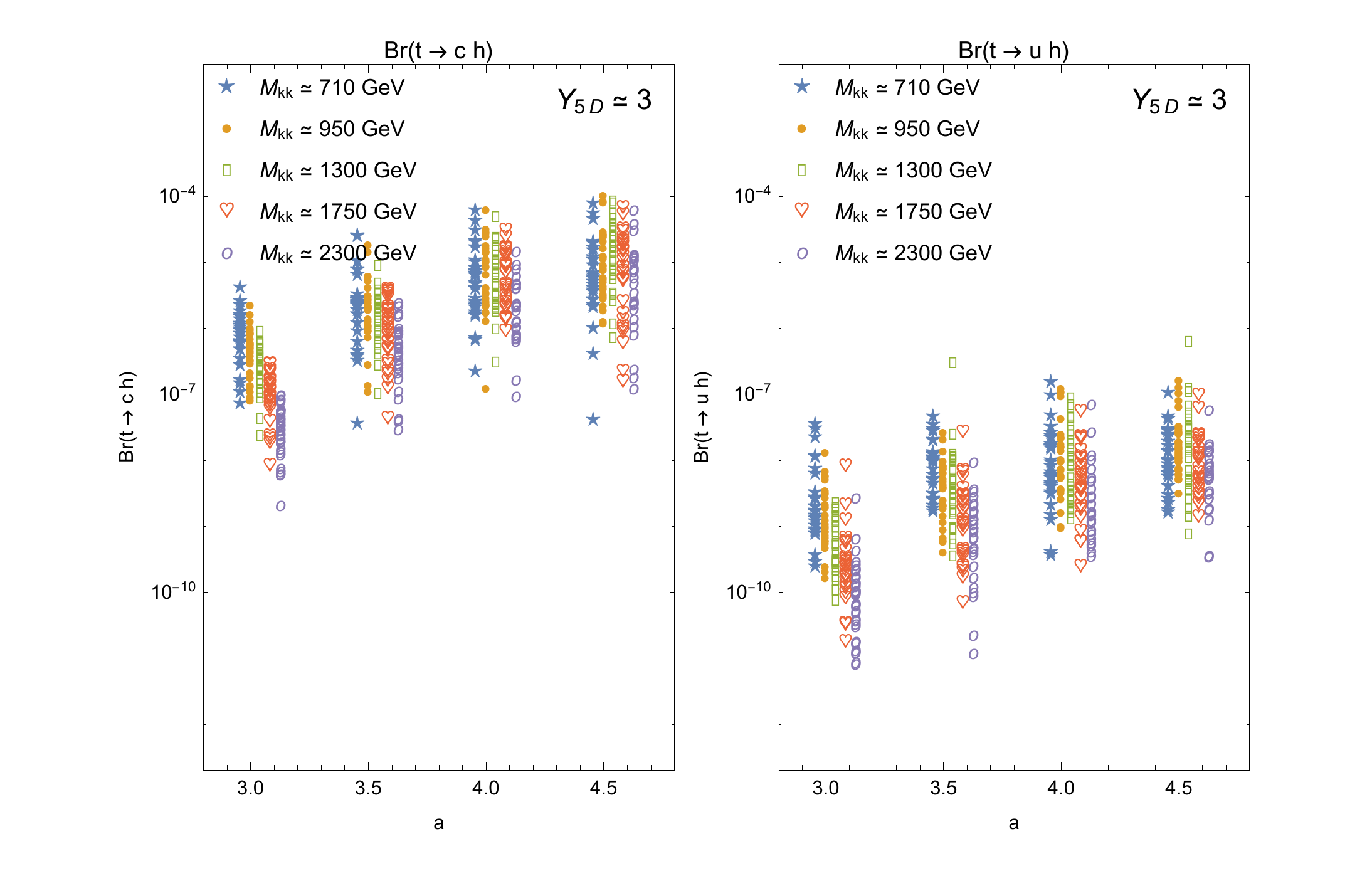}
	\includegraphics[height=10.cm, width=17.5cm]{\plots/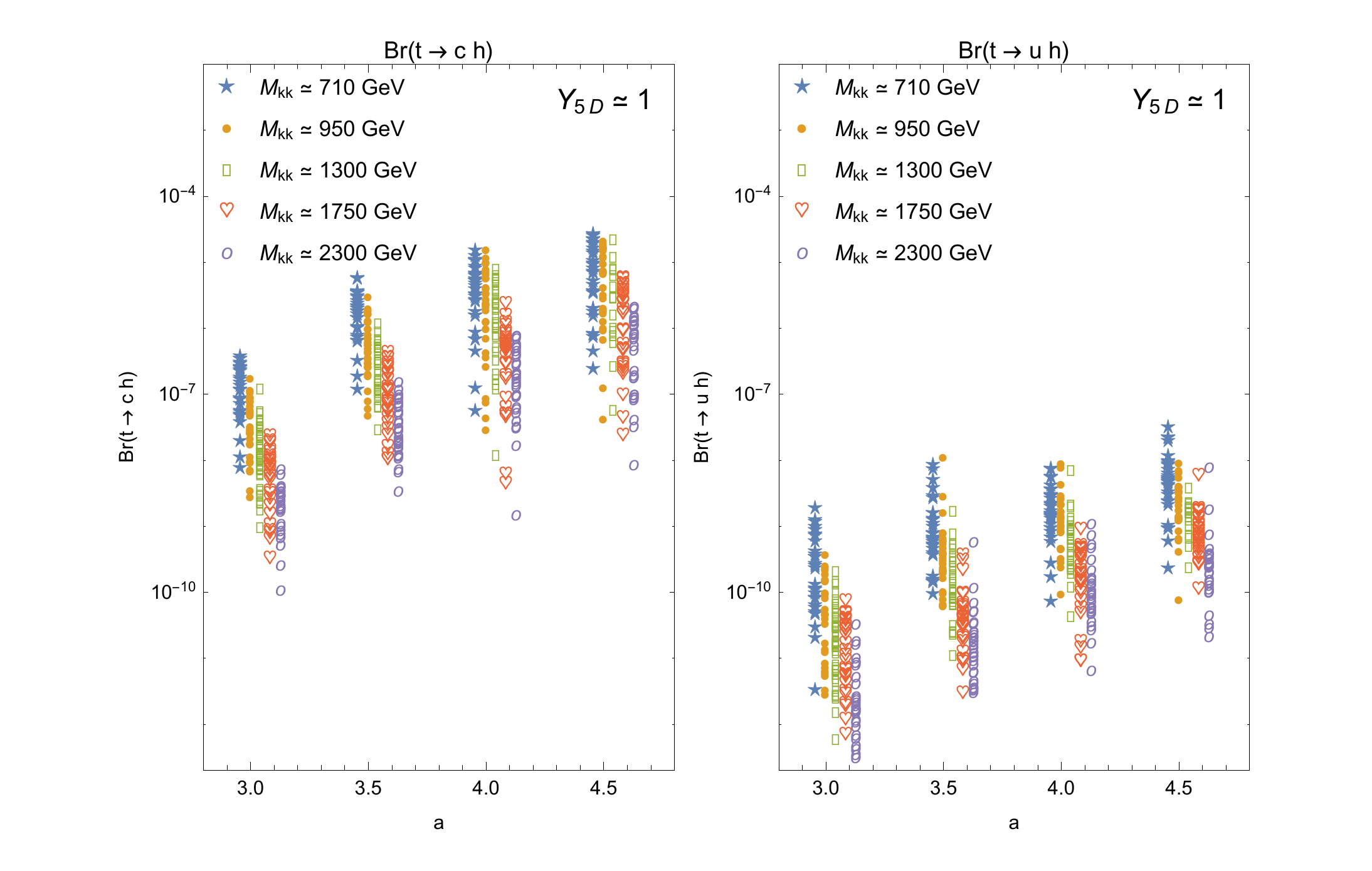} 	
 \end{center}
\caption{Branching ratios for the tree-level decays $t\to ch$ (left
  panels), $t\to uh$ (right panels) for four values of the Higgs
  localization parameter, $a\simeq \{3, 3.5, 4, 4.5\}$. Two different
  5D Yukawa average values are used, $Y^{5D}=3$ in the
  upper panels and $Y^{5D}=1$ in the lower panels, and five different
  KK gluon mass scales between $710$ GeV and $2300$ GeV set the KK
  scale. It is important to note that within the 
  $Y^{5D}=1$ case, the (33) entry in the up-Yukawa matrix must still
  be about $\sim 3$ in order to reproduce the top quark mass.}  
  \label{fig:tqh}
\end{figure}

The results of this calculation are shown in Fig. \ref{fig:tqh}, where
four values of $a$ have been used as explained earlier. Since we have
also included five different KK masses, we present the results for the
same $a$ but different KK mass, slightly shifted in the $a$ scale. 

The branching ratio is expected to depend inversely on the KK
scale and directly related to the Yukawa couplings, roughly as
$\displaystyle {\cal B} (t \rightarrow q h)\sim \frac{Y_{5D}^4}{M_{KK}^4}\frac{m_q}{v} $ \cite{Azatov:2009na}.
We see that the dependance on the KK mass more or less follows that
trend, although for larger 5D Yukawa couplings ($Y^{5D}\sim 3$) the dependence seems 
milder. This large Yukawa case might be on the verge of validity for 
our perturbative calculations since the full flavor effects accelerate
the appearance of strong coupling effects.
Note also that in the case ($Y^{5D}\sim 1$), all Yukawa couplings
are order $\sim 1$ ({\it i.e.} safer), except for the (33) entry which must still be of
order $\sim 3$ in order to generate the top quark mass. This means
that even in the smaller Yukawa case there is a somewhat larger flavor
effect. Nevertheless no cumulative flavor family enhancements are
present in this case, making it safe in terms of perturbativity.

Notice also that as the Higgs becomes more and
more localized towards the IR brane, \ie \, larger values of $a$, the
branching ratio increases. This is due to an enhancement of the
overlap integrals as the more IR localized Higgs field couples more
strongly with the fermion fields \footnote{These results are in
  agreement with our previous results \cite{Frank:2013un,
    Frank:2013qma, Frank:2015zwd}.}. Unfortunately, larger Higgs 
couplings also lead to stronger bounds from precision electroweak
constraints as well as from the Higgs production and decay phenomenology. The
threshold is not clear cut, but for the larger values of $a$
considered, one should expect that only the larger KK scales might be
safe from constraints. Thus the expected size of the branching ratio for the
$t \to c h$ decays should be somewhere around ($10^{-5}$ - $10^{-7}$), and
for  $t \to u h$ somewhere around  ($10^{-6}$ - $10^{-8}$).
These last two ranges are consistent with expectations as one would
expect a relative strength governed by (very roughly) $m_u/m_c$.


We now proceed to analyze the other  tree-level top quark FCNC decay,
to the $Z$ boson, coming from terms in the Lagrangian, 
\bea
{\cal L} \ni \left( g^u_{U_{L, R}}{\bar t}_R \gamma_{\mu} c_R
+g^u_{Q_{L, R}} {\bar t}_L \gamma_{\mu} c_L \right) Z^{\mu}. 
\eea
In order to obtain the couplings in the mass basis, we must
calculate first the following overlap integrals in the gauge basis, among left handed and right
handed KK fermions and the $Z$ boson wavefunction,   
\begin{equation}\label{gZ1}
(g^u_{Q_{L, R}})_{ij} = \frac{g^{5D}_{L,R}}{\sqrt{k}} \int_0^{y_1} dy e^{-3A(y)} f_Z(y) Q^{m,i}_{L, R}(y)Q^{n,j}_{L, R}(y),
\end{equation}
\begin{equation}\label{gZ2}
(g^u_{U_{L, R}})_{ij} = \frac{g^{5D}_{L,R}}{\sqrt{k}} \int_0^{y_1} dy e^{-3A(y)} f_Z(y) U^{m,i}_{L, R}(y)U^{n,j}_{L, R}(y),
\end{equation}
where as usual $Q^{n,j}_{L, R}(y)$ stands for the $SU(2)$ doublets and
$U^{n,j}_{L, R}(y)$ for $SU(2)$ singlets. We then need to 
perform a rotation on the quark fields to transform to the physical mass
basis. This rotation will produce tree-level flavor violating couplings of the $Z$ boson.
The profile $f_Z(y)$ is the solution to the following differential equation
\begin{equation}
\partial_y\left( e^{-2 A(y)} \partial_y f_Z(y)\right) - M_A^2 f_Z(y) + m^2 f_Z = 0
\end{equation}
with $M_Z(y) = \displaystyle {g^{5D} \over 2 \cos \theta_W} h(y)
e^{-A(y)}$ being the $y$ dependent bulk mass of the field.  
The $g^{5D}_{L,R}$ coupling are given by
$$
g^{5D}_{L} =  {g^{5D}\over \cos \theta_W } \left(T_3 - Q_q \sin^2 \theta_W  \right)
$$
$$
g^{5D}_{R} = {g^{5D}\over \cos \theta_W} Q_q \sin^2 \theta_W ,
$$
with $Q_q$  the charge of the quark, (here ${2 \over 3}$),
$\theta_W$ the Weinberg angle and $T_3 = {1\over 2}$. 
Once in the mass basis we extract the flavor violating couplings $(g^u_L)_{qt}$ and
$(g^u_R)_{qt}$ between right handed and left handed quarks and the $Z$
boson and with these, the flavor violating branching ratio is given by \cite{Casagrande:2008hr}
 \begin{equation}
{\cal B} (t \rightarrow q Z) =
{2 \left(1-r_Z\right)^2 \left(2 r_Z+1\right) \over
  \left(1-r_W\right)^2 \left(2 r_W+1\right)}
\left(\left|(g^u_L)_{\text{qt}}\right|^2+\left|(g^u_R)_{\text{qt}}\right|^2
- {12 \sqrt{r_q} r_Z \over \left(1-r_Z\right) \left(2 r_Z+1\right) }
{\text Re} \left[ (g^u_L)^*_{\text{qt}} (g^u_R)_{\text{qt}} \right]\right).
\end{equation}
\begin{figure}[t]
\vspace{-1cm}
\center
\begin{center}
	\includegraphics[height=10.cm, width=17.5cm]{\plots/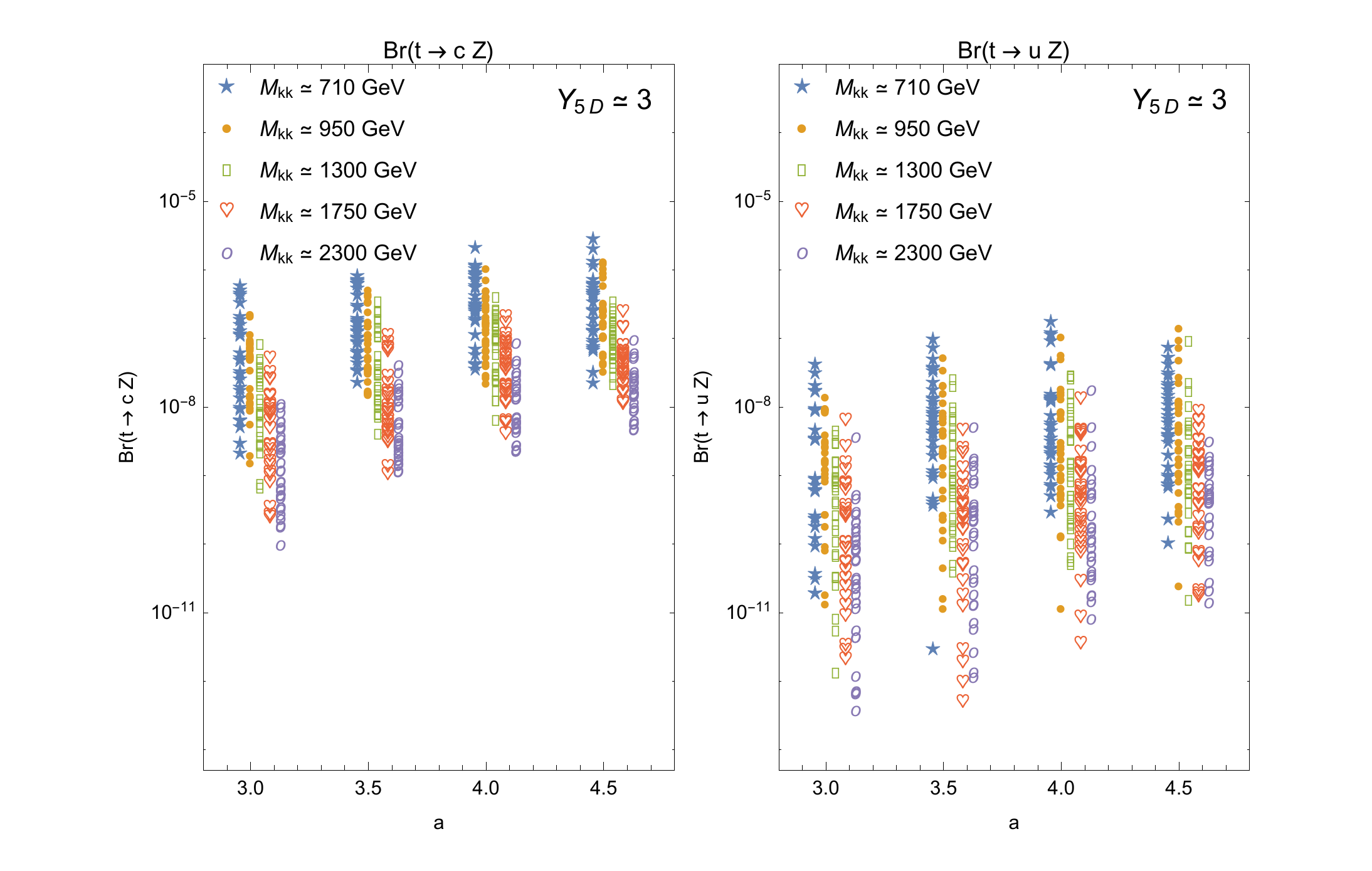} \hspace{.5cm}
	\includegraphics[height=10.cm, width=17.5cm]{\plots/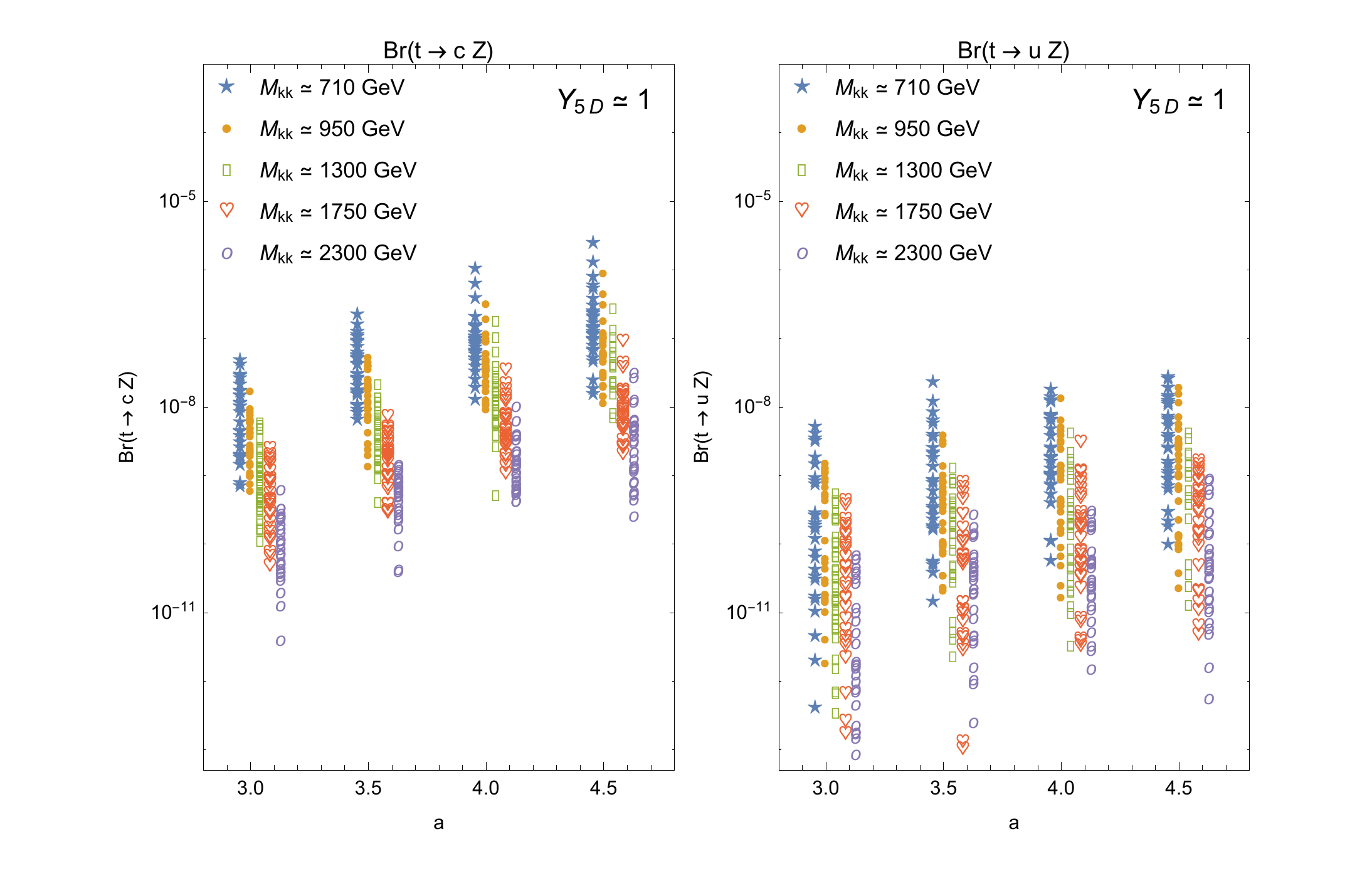} 	
 \end{center}
\caption{Branching ratios for the tree-level decays $t \to c Z$ (left
  panels), $t \to u Z$ (right panels) for the Higgs
  localization parameter, $a\in \{a_{\text min}, a_{\text min}+0.5,
  a_{\text min}+1, a_{\text min}+1.5
  \}$, where
  $a_{\text min}\simeq 3$. We have presented our results for two
  different scales of 5D Yukawa couplings, $Y^{5D}=1, 3$ ($Y^{5D}=3$
  in the upper panels, and for $Y^{5D}=1$ in the lower panels), and
  five different KK gluon mass scales between $710$ GeV to $2300$
  GeV. 
}
\label{fig:tqZ}
\end{figure}
 Our results are presented in  Fig. \ref{fig:tqZ}, where again the
 branching ratio is expected to scale as $\displaystyle {\cal B} (t \rightarrow q
 Z)\propto \frac{Y^4}{M_{KK}^4} \frac{m_q}{v}$, although in this case there will
 not be a flavor cumulative effect as in the Higgs case \cite{Azatov:2009na}, and so the
 overall effect is expected to be smaller than the flavor violating top decay into a Higgs. 
We observe that the $Y^{5D}=1$ and $Y^{5D}=3$ graphs show very similar branching ratios
for the same values of $a$ and given KK scale. This is consistent with
the fact that even in the $Y^{5D}=1$ case, there is still one large
Yukawa entry, ($Y^{5D}_{33}\sim3$). That single term must dominate the
overall effect, so that both Yukawa scenarios give similar results (i.e. the
rest of 5D Yukawa couplings do not seem to add up constructively as
they did in the flavor violating Higgs couplings case).

The KK scale dependence is as expected, i.e. the suppression due to
larger KK masses scales down consistently, so that masses 3 times heavier, produce a branching ratio about
2 orders of magnitude smaller. Finally the flavor dependence between
charm and up quark also seems consistent as there should be (very roughly) some
factor of $m_u/m_c$ between the flavor violating decay branching of $t\to cZ$ over
that of $t\to uZ$.

\subsection{Flavor-violating radiative decays of the top quark: $t \to q \gamma$ and $t \to q g$}
\label{subsec:analytic}
\begin{figure}[t]
\center
\begin{center}
$\begin{array}{cc}
	\includegraphics[height=5.cm]{\diagrams/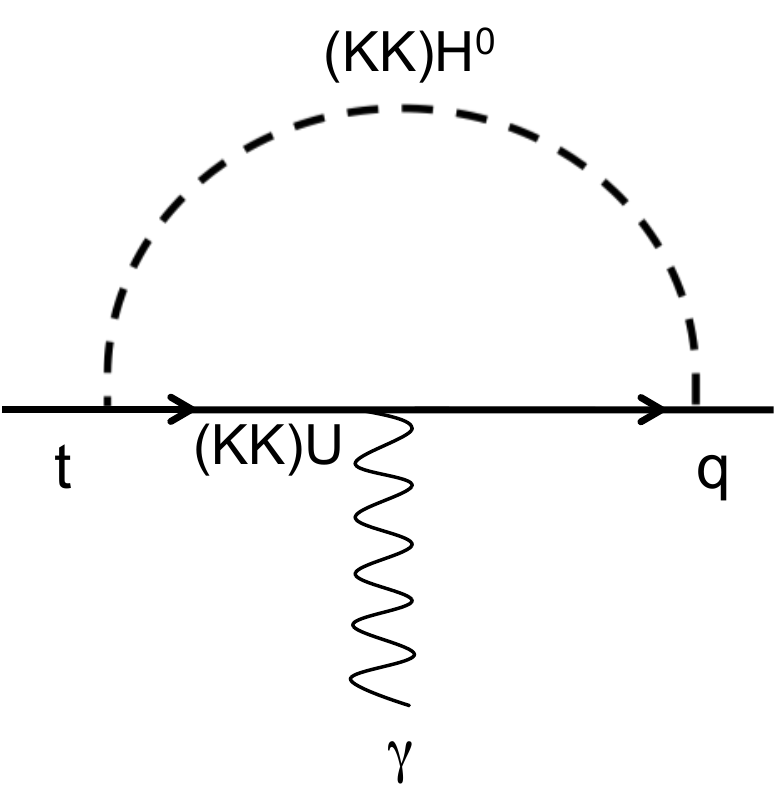} & \includegraphics[height=5.cm]{\diagrams/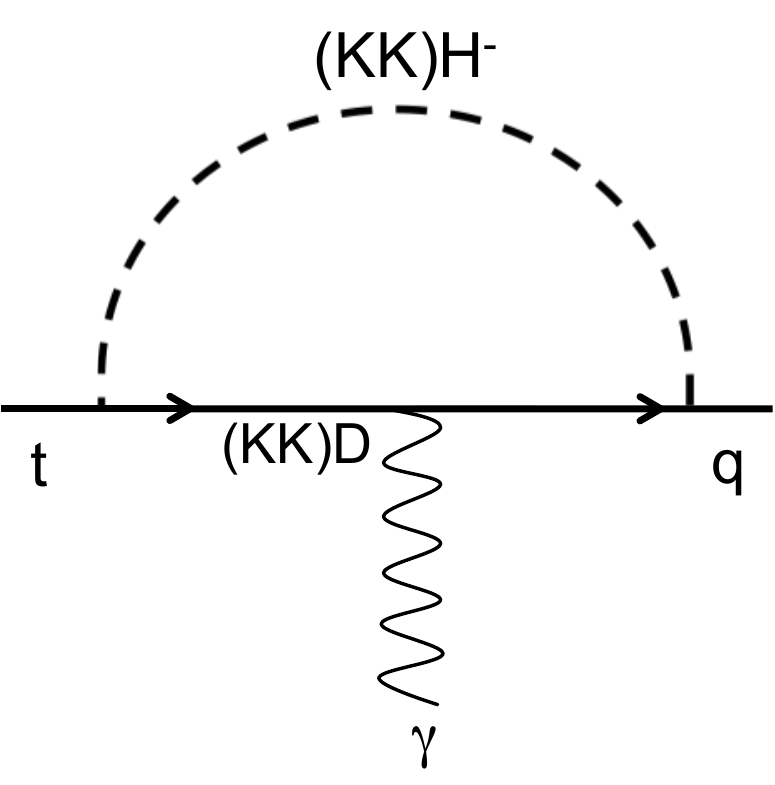} \\
	\includegraphics[height=5.cm]{\diagrams/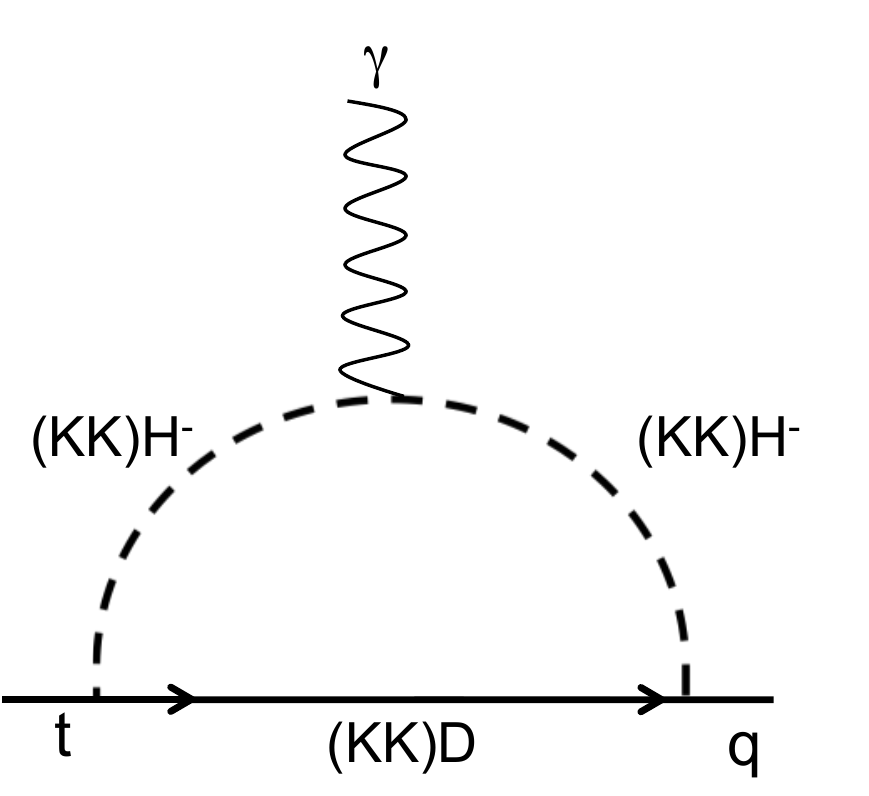} & \includegraphics[height=4cm]{\diagrams/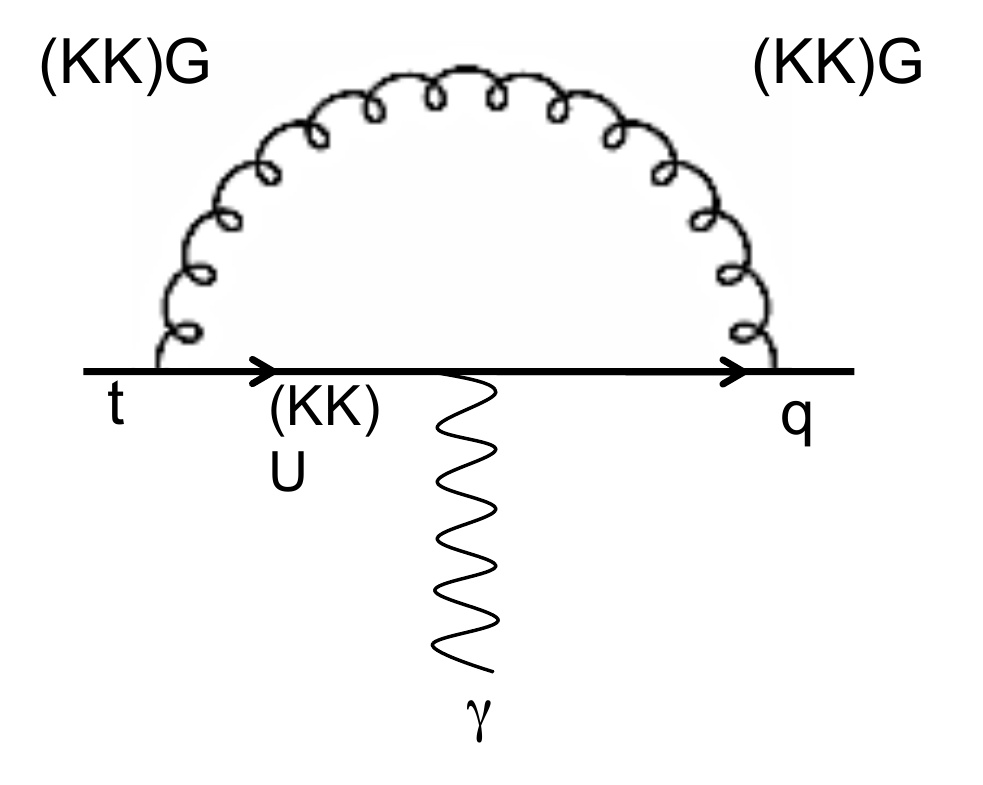}
	\end{array}$	
 \end{center}
\caption{Feynman Diagrams for the decay $t \to c \gamma$ in warped
  extra dimensional models. The label (KK) implies summation over the
  zero and the KK modes, while KK only indicates that only KK modes
  contribute.} 
   \label{fig:diagrams1}
\end{figure}

\begin{figure}[t]
\center
\begin{center}
$\begin{array}{cc}
	\includegraphics[height=5.cm]{\diagrams/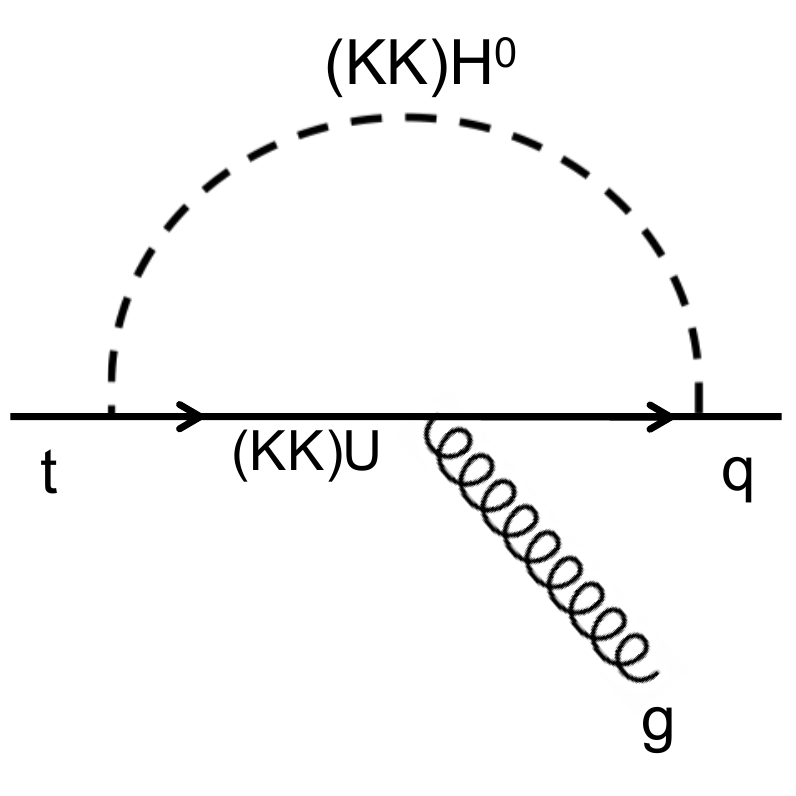} & \includegraphics[height=5.cm]{\diagrams/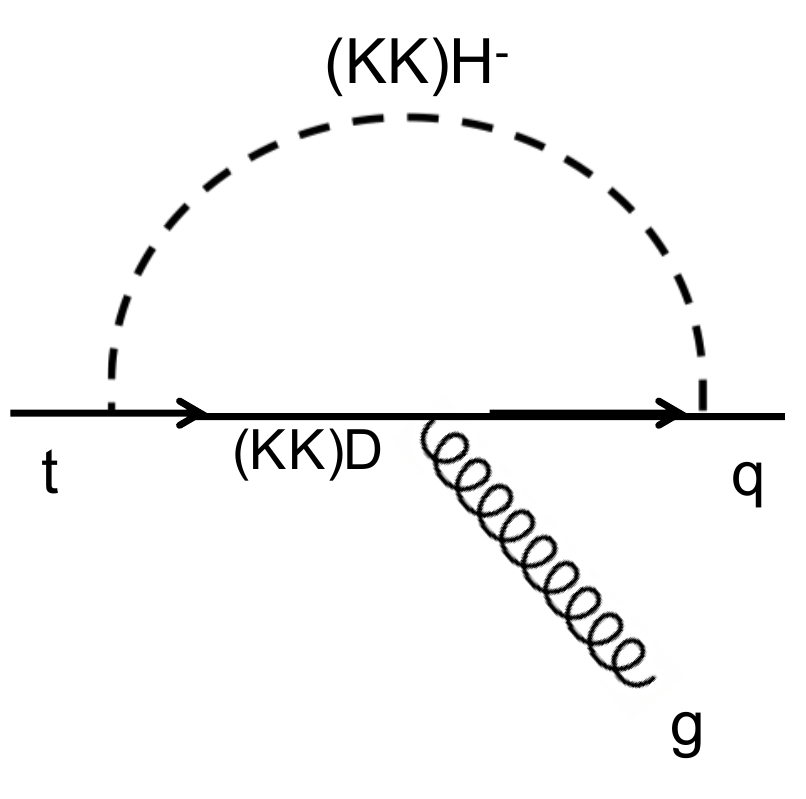} \\
	\includegraphics[height=5.cm]{\diagrams/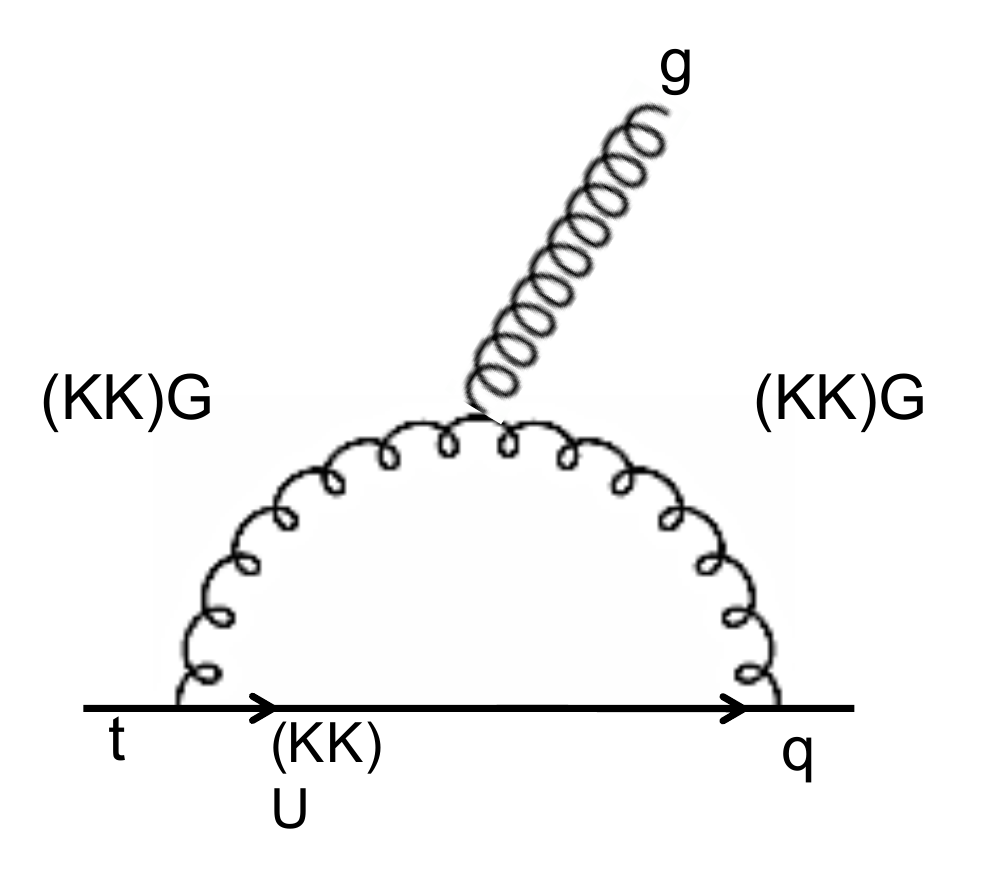} & \includegraphics[height=4cm]{\diagrams/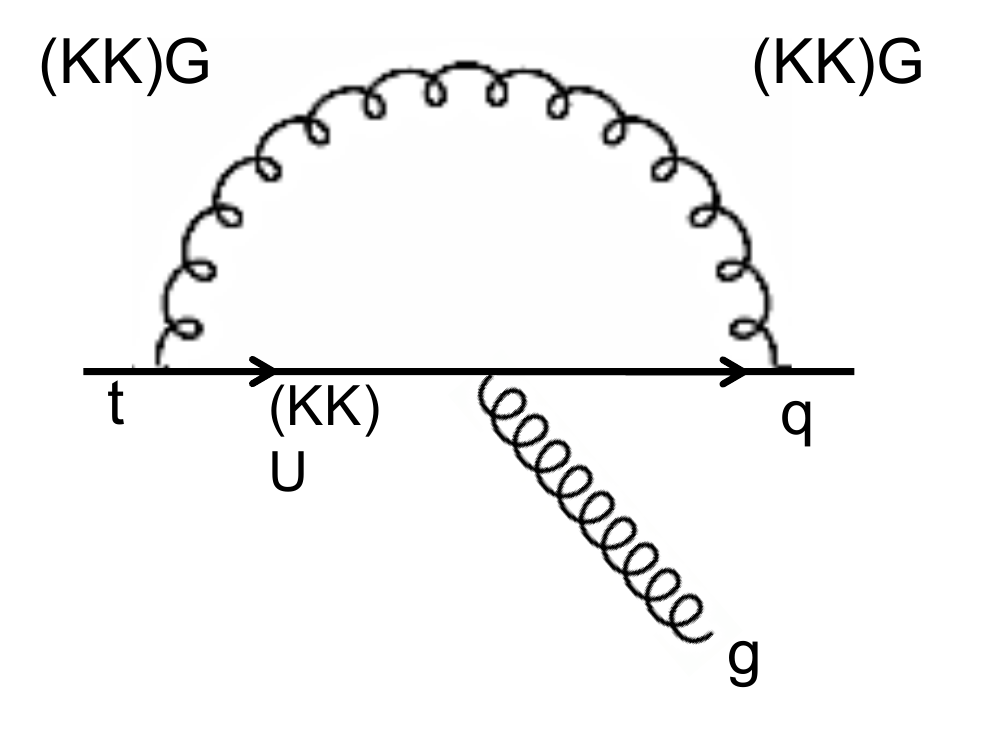}
	\end{array}$	
 \end{center}
\caption{Feynman Diagrams for the decay $t \to c g$ in warped extra
  dimensional models. The label (KK) implies summation over the zero
  and the KK modes, while KK only indicates that only KK modes
  contribute.} 
   \label{fig:diagrams2}
\end{figure}

The SM predicted values for the ${\cal B}(t \to c \gamma)$ and ${\cal
  B}(t \to c g)$ are far from the LHC sensitivity. As the expected
sensitivity to reach the branching ratio of $t \to c g, t \to c\gamma$
at the LHC is $10^{-5} - 10^{-6}$, observing these decays would also indicate
an important evidence of physics beyond the SM.  These flavor violating interactions
are described by the following effective Lagrangian 
\begin{eqnarray} 
{\cal L}_{FCNC}&=& i \sum\limits_{q} \{ {\bar q}\left (C_{8 g}^L P_L +
C_{8 g}^R P_R \right ){\sigma_{\mu \nu} q^\nu} t T^a G^{a\,\mu} \}
+h.c. \nonumber \\ 
&+& i \sum\limits_{q} \{ {\bar q}\left (C_{7 \gamma}^L P_L + C_{7
  \gamma}^R P_R \right ){ \sigma_{\mu \nu} q^\nu} t  A^\mu \}+ h.c.\, . 
\end{eqnarray}
The Feynman diagrams for this processes are shown in
Fig.~\ref{fig:diagrams1} for ($t \to c \gamma$), and in
Fig.~\ref{fig:diagrams2} for ($t \to cg$), and the analytical
expressions needed for the calculation of the branching ratio are given in the Appendix 
(Section \ref{sec:appendix}).  

The scenario contains a tower of physical neutral scalars coming from
the real component of the 5D Higgs doublet. There is also a tower of
physical charged scalars, which contain a mixture of the charged
component of the 5D Higgs doublet and the fifth component of charged
5D gauge  boson (although for simplicity we will refer to them as charged KK
Higgses). Finally there is a further tower of neutral CP-odd
scalars, containing a mixture of the CP-odd component of the 5D Higgs
doublet and the fifth component of neutral 5D gauge boson. The
scenario contains also towers of charged and neutral Goldstone bosons, orthogonal
admixtures of the previous 5D Higgs and gauge boson degrees of freedom
(see for example \cite{Cabrer:2011fb} for details).

In the calculation of loop effects we consider the truncated tower of
the first 3 KK neutral Higgs modes (CP even and odd) as well as the first 3 KK 
charged Higgs modes, each coupled with the first 3 KK mode fermions
(\ie \, four modes in total, including the 
zero mode). The neutral Higgs couplings are the same as the ones
given in Eq. (\ref{yumatrix}) but we need to compute also the
Yukawa couplings between fermions and KK Higgs modes 
\begin{equation}\label{YKKHiggs}
(Y_u)_{ij}^{H_{k}} = \frac{Y^{u, 5D}_{ij}}{\sqrt{k}} \int_0^{y_1}
  dy e^{-4A(y)} h^{H_{k}}(y) Q^{m,i}_{L, R}(y)U^{n,j}_{R, L}(y), 
\end{equation}
constructing the corresponding ($21 \times 21$) matrices in the same
basis. These must then be rotated as  in Section \ref{sec:pheno} to
be transformed into the mass basis.

In the case of the charged KK Higgs couplings we have the following
coupling matrices written in the same basis as the rest of Yukawa matrices
\begin{equation}
{\bf Y}^{H_k^{+}LR} = \left(\begin{array}{ccc} 
(y^{H^+}_{u})_{3\times3}      &  (0)_{3\times 3N}   & (Y^{H^{+}qU})_{3\times 3N}\\
 (Y^{H^{+}Qu})_{3N \times3}      & (0)_{3N\times  3N} & (Y^{H^+Q_LU_R})_{3N \times 3N}\\
(0)_{3N \times3}      &     (Y^{H^+D_LQ_R})_{3N\times  3N} &  (0)_{3N\times 3N}
  \end{array}\right),
  \label{yplusLR}
\end{equation}
and
\begin{equation}
{\bf Y}^{H_k^{+}RL} = \left(\begin{array}{ccc} 
(y^{H^+}_{d})_{3\times3}      &  (0)_{3\times 3N}   & (Y^{H^+qD})_{3\times 3N}\\
 (Y^{H^+Qd})_{3N \times3}      & (0)_{3N\times  3N} & (Y^{H^+Q_LD_R})_{3N \times 3N}\\
(0)_{3N \times3}      &     (Y^{H^+U_LQ_R})_{3N\times  3N} &  (0)_{3N\times 3N}
\end{array}\right),\label{yplusRL}
\end{equation}
where the submatrices are obtained by evaluating the overlap integrals 
\bea\label{YYYplusLR}
y_{u}^{H^+} &=& \frac{(Y^{5D}_u)_{ij}}{\sqrt{k}} \int_0^{y_1} dy e^{-4A(y)} \xi^k(y) q^{0,i}_L(y)u^{0,j}_R(y)\\  
Y^{H^+qU}&=& \frac{(Y_u^{5D})_{ij}}{\sqrt{k}} \int_0^{y_1} dy e^{-4A(y)} \xi^k(y) q^{0,i}_{L}(y)U^{n,j}_{R}(y) \\
Y^{H^+Qu} &=& \frac{(Y_u^{5D})_{ij}}{\sqrt{k}} \int_0^{y_1} dy e^{-4A(y)} \xi^k(y) Q^{m,i}_{L}(y)u^{0,j}_{R}(y)\\
Y^{H^+Q_LU_R} &=& \frac{(Y_u^{5D})_{ij}}{\sqrt{k}} \int_0^{y_1} dy e^{-4A(y)} \xi^k(y) Q^{m,i}_{L}(y)U^{n,j}_{R}(y) \\
Y^{H^+Q_RD_L} &=& \frac{(Y_d^{5D})^*_{ij}}{\sqrt{k}} \int_0^{y_1} dy e^{-4A(y)} \xi^k(y) Q^{m,i}_{R}(y)D^{n,j}_{L}(y)\, ,
\eea
and 
\bea\label{YYYplusRL}
y_{d}^{H^+} &=& \frac{(Y^{5D}_d)_{ij}}{\sqrt{k}} \int_0^{y_1} dy e^{-4A(y)} \xi^k(y) q^{0,i}_L(y)d^{0,j}_R(y)\\  
Y^{H^+qD}&=& \frac{(Y_d^{5D})_{ij}}{\sqrt{k}} \int_0^{y_1} dy e^{-4A(y)} \xi^k(y) q^{0,i}_{L}(y)D^{n,j}_{R}(y) \\
Y^{H^+Qd} &=& \frac{(Y_d^{5D})_{ij}}{\sqrt{k}} \int_0^{y_1} dy e^{-4A(y)} \xi^k(y) Q^{m,i}_{L}(y)d^{0,j}_{R}(y)\\
Y^{H^+Q_LD_R} &=& \frac{(Y_d^{5D})_{ij}}{\sqrt{k}} \int_0^{y_1} dy e^{-4A(y)} \xi^k(y) Q^{m,i}_{L}(y)D^{n,j}_{R}(y) \\
Y^{H^+Q_RU_L} &=& \frac{(Y_u^{5D})^*_{ij}}{\sqrt{k}} \int_0^{y_1} dy e^{-4A(y)} \xi^k(y) Q^{m,i}_{R}(y)U^{n,j}_{L}(y)\,.
\eea
We then need to transform into the physical mass basis, by appropriately rotating from
the left and the right using the two different rotation matrices
required to diagonalize the up mass matrix and the down mass
matrix. With this procedure, the first matrix will generate the
interactions $H^+d_Lu_R$ and the second one will produce the terms $H^+u_Ld_R$.

For the KK gluon contributions to the loop, we compute the kinetic overlap integrals
similar to the ones for the $Z$ boson couplings, Eqs. (\ref{gZ1}) and
(\ref{gZ2}), but with the KK gluon fields replacing the $Z$ fields. 
\bea
\chi_L^{G^kQ^iQ^j}  
&=& \frac{g^{5D}}{\sqrt{k}} \int_0^{y_1} dy e^{-3A(y)} f_G^k(y) {Q}^{i}_{L}(y) 
{Q}^{j}_{L}(y),\\    
\chi_L^{G^kU^iU^j}  
&=& \frac{g^{5D}}{\sqrt{k}} \int_0^{y_1} dy e^{-3A(y)} f_G^k(y) {U}^{i}_{L}(y) 
{U}^{j}_{L}(y),\\  
\chi_R^{G^kQ^iQ^j}  
&=& \frac{g^{5D}}{\sqrt{k}} \int_0^{y_1} dy e^{-3A(y)} f_G^k(y) {Q}^{i}_{R}(y) 
{Q}^{j}_{R}(y),\\    
\chi_R^{G^kU^iU^j}  
&=& \frac{g^{5D}}{\sqrt{k}} \int_0^{y_1} dy e^{-3A(y)} f_G^k(y) {U}^{i}_{R}(y) 
{U}^{j}_{R}(y),
\eea
where $g^{5D} = \sqrt{4 \pi \alpha_S y_1}$.

\begin{figure}[t]
\vspace{-1cm}
\center
\begin{center}
	\includegraphics[height=10.cm, width=17.5cm]{\plots/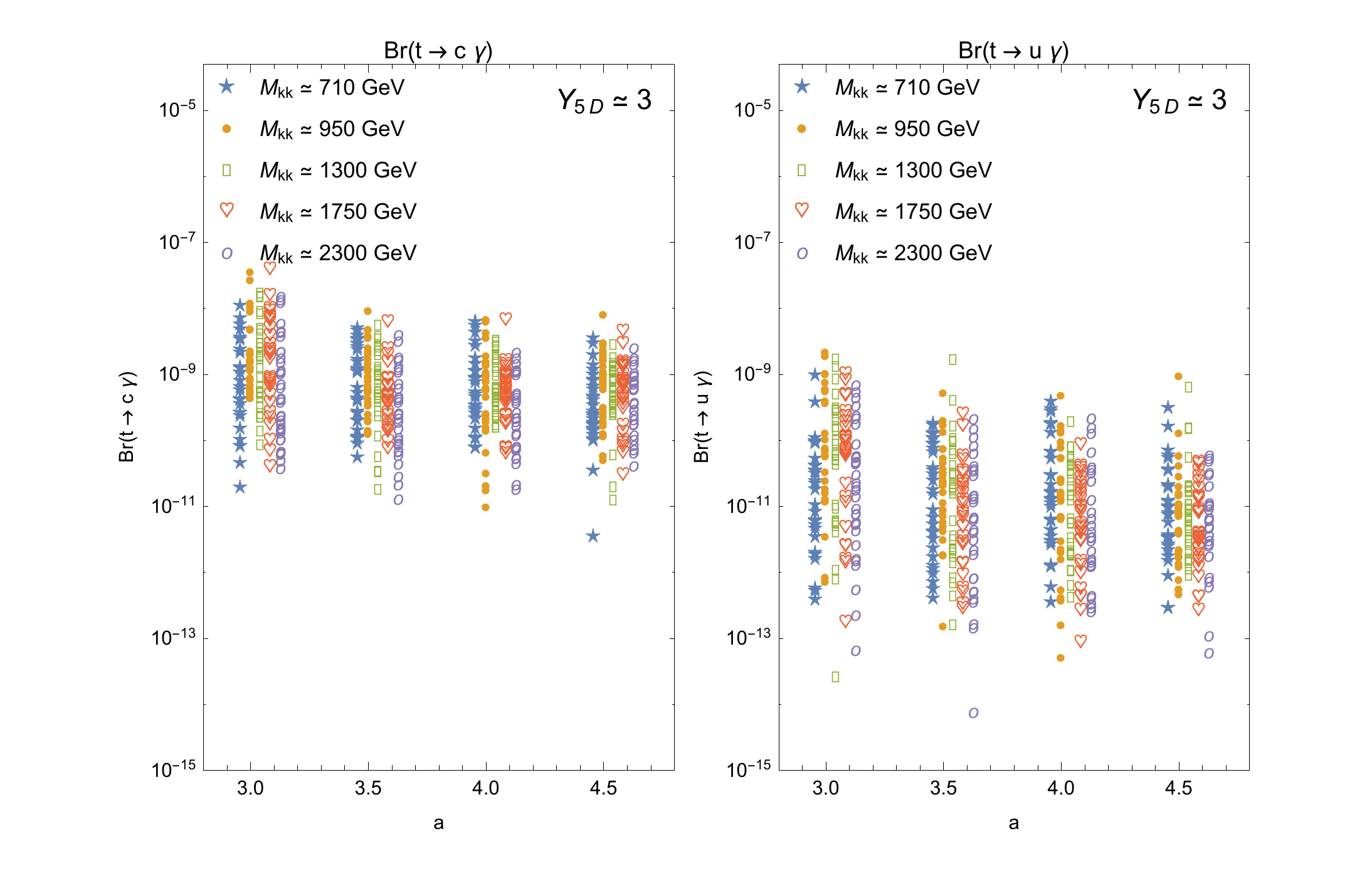} \hspace{.5cm}\vspace{-.5cm}
	\includegraphics[height=10.cm, width=17.5cm]{\plots/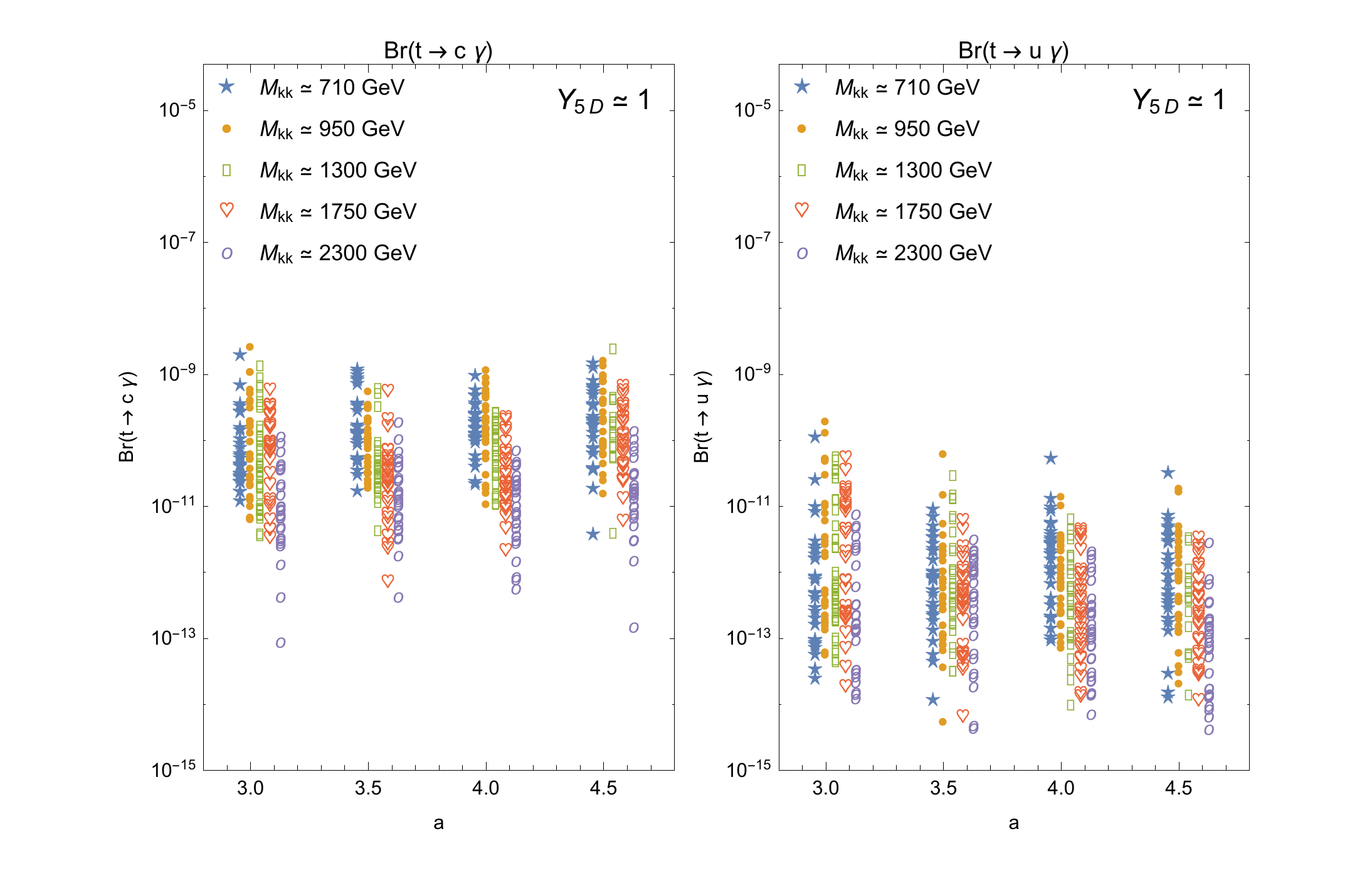} 	
 \end{center}\vspace{-.5cm}
\caption{Branching ratios for the  decays $t\to c\gamma$ (left panels)
  and $t \to u \gamma$ (right panels) for the the Higgs localization
  parameter $a$. We have presented our results for two different
  scales of 5D Yukawa couplings, $Y^{5D}=1, 3$ ($Y^{5D}=3$ in the
  upper panels, and for $Y^{5D}=1$ in the lower panels), and five
  different KK gluon mass scales between $710$ GeV to $2300$ GeV. 
 }  \label{fig:tqgam}\vspace{-.5cm}
\end{figure}

\begin{figure}[t]
\vspace{-1cm}
\center
\begin{center}
	\includegraphics[height=10.cm, width=17.5cm]{\plots/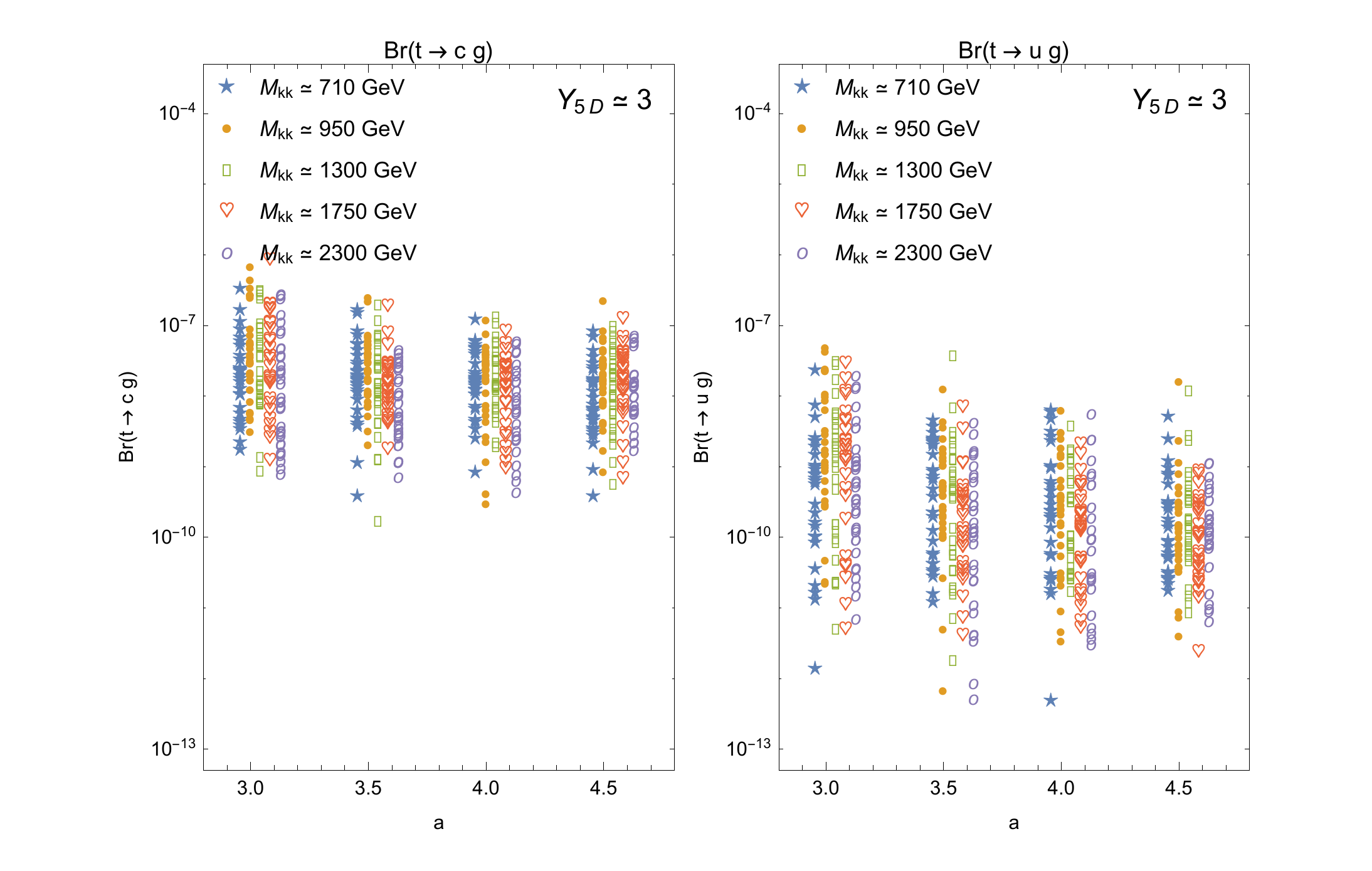} \hspace{.5cm}
	\includegraphics[height=10.cm, width=17.5cm]{\plots/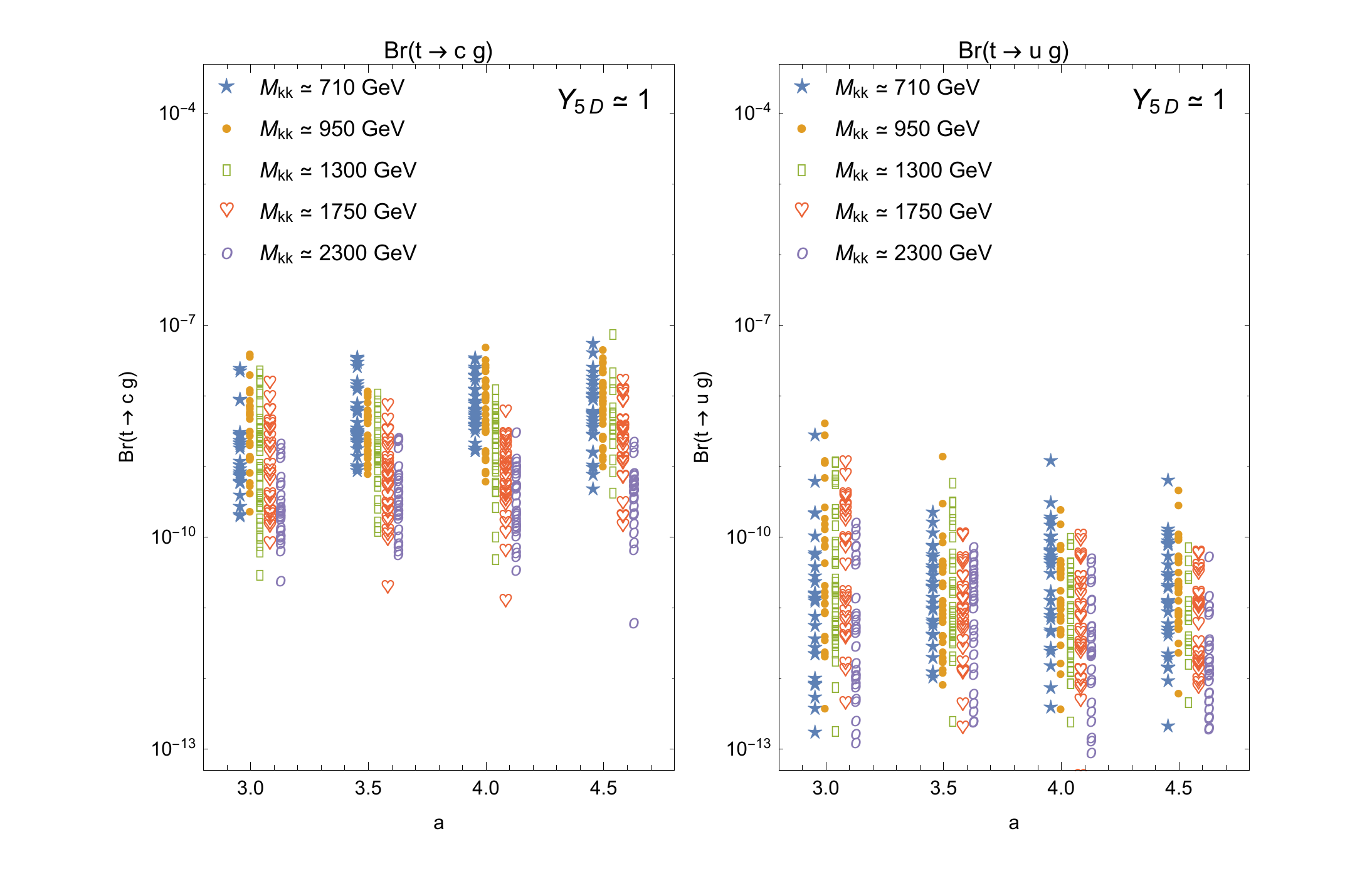} 	
 \end{center}
\caption{Branching ratios for the  decay $t\to cg $ (left panels) and $ t \to ug$ (right panels) for the the Higgs localization parameter $a$. We have presented our results for two different scales of 5D Yukawa couplings, $Y^{5D}=1, 3$ ($Y^{5D}=3$ in the upper panels, and for $Y^{5D}=1$ in the lower panels), and five different KK gluon mass scales between $710$ GeV to $2300$ GeV.
}  
\label{fig:tqg}
\end{figure}

With these matrix elements we can compute the $t \rightarrow q \gamma$
and $t \to q g$ decay rates given in \ref{sec:appendix}. The results
of the numerical computations are presented in Figs.  \ref{fig:tqgam}
and \ref{fig:tqg}.

We observe that the decay branching ratios are at most $10^{-7}$ for $t\to cg$
and $10^{-8}$ for $t\to c\gamma$, and these obtained with very light KK gluon
masses of about $700$ GeV (KK Higgs are heavier).
In any case, the sensitivity at the LHC for the $t \to
q g$ decay is expected to be at about $\sim 10^{-5}$ which renders
these results still too small to be observed.

We also observe that the expected behavior for heavier and heavier
masses is not completely apparent, specially for large Yukawa couplings and large values of the Higgs
parameter $a$. First of all, as we mentioned in Sec. \ref{subsec:tree} the validity of our
perturbative approach becomes questionable for larger values of Yukawa
couplings and of the $a$ parameter, which, when including the flavor family effects, could start to fail for
\ie \, $Y^{5D} \simeq 3$ and for IR localized Higgs field, $a > 4$
(\ie \, for those parameters one should include another full KK level of fields to improve the
situation assuming that we have not reached the strong coupling limit).
In the loop calculations for  $t \rightarrow q \gamma$  and $t \to q g$ this
effect might be partly at play in the most extreme regions of the parameter space
 shown and one should therefore look at the results presented
for large Yukawa couplings and large values of $a$ as being close to the limit
of validity of our approach. 

Nevertheless, there seems to be also a specific characteristic
behavior of the scenario at play here.
After scanning the parameter space, in all loop evaluations, it appears that the Higgs
  and KK Higgs contributions to the loop are larger than the KK gluon
  contributions by one order of magnitude.
  This should be in part responsible for the fact that the loop
  processes do not appear very sensitive to the KK mass scale. The
  reason for this is that the mass scale indicated on the plots
  represents the lowest KK gluon mass, while the computed KK Higgs
  masses are  always larger and less sensitive to changes in the KK volume as
  the KK gluons, in the first portion of parameter range. In particular, when
  the lightest KK gluon is $710$ GeV, the lightest KK Higgs is $\sim 1300$ GeV,
  while when the KK gluon mass is $950$ GeV, the lightest KK Higgs remains at
  $\sim 1300$ GeV. When the KK gluon mass is increased to $1300$ GeV, the
  lowest KK Higgs mass becomes $1500$ GeV, and from then on, the increases in masses
  remain relatively similar.  This means that the
 effect expected from changing the  KK scale is milder since the lightest KK
  Higgs have similar masses despite increasingly
  heavier KK gluons, thus obscuring the mass scale dependence (as
  the loops are dominated by the KK Higgs).

\section{Summary and Conclusion}
\label{sec:conclusion}
We presented a comprehensive analysis of FCNC decays of the top quark,
both at tree-level and  one-loop, in the context of a general warped
extra dimensional space scenario, with all the fields in the bulk,
such that a low KK scale is allowed without violating electroweak 
precision constraints.  

We first constructed a complete KK tower of scalar, fermion and gauge
boson states, consistent with the experimental data. We imposed
constraints on the masses and mixing of zero-mode fermions so as to be
consistent with the known quark masses and the CKM mixing matrix, and
therefore limiting the possible values of the fermion localization
parameters $c_i$. 
We used the fermion and scalar profiles to
analyze the FCNC branching ratios of the top quark, both at tree and
one-loop level, as functions of the Higgs localization parameter $a$
(bulk localized Higgs) and for various KK scales, allowing the 5D
Yukawa couplings for the quarks to be of order $Y^{5D}=1$ and $Y^{5D}=3$. We
performed all our calculations in the mass eigenstate basis, where we
can take into account quark inter-generational mixing.  This involved
diagonalizing the $21 \times 21$ dimensional fermion matrices and rotating
various  $21 \times 21$ Yukawa coupling matrices and KK gauge-fermion-fermion matrices. 

The most promising of these decays is the tree-level $t \to c h$,
whose branching ratio could reach ${\cal O} (10^{-5})$ and thus become observable
at the LHC @13 TeV. We note that while the branching ratio is slightly
higher for $Y^{5D}=3$ (which might reach the non-perturbativity limit,
particularly for highly IR localized Higgs), it is still of the
same order of magnitude as for $Y^{5D}=1$ (in this case, most of the
Yukawa couplings are ${\cal O}(1)$, but there must always be a larger than
normal Yukawa entry, $Y^{5D}_{33} \sim 3$, in order to reproduce the top
quark mass).
The tree-level decay $t \to c Z$ is, as expected, a
couple of orders of magnitude smaller, as it is driven by kinetic
mixing rather than Yukawa couplings. At one-loop level the branching
ratio for the decay $t \to c \gamma$ can be at most of ${\cal
  O}(10^{-8})$, and the one for $t \to c g$ an order of magnitude
larger. Both of these
contributions are dominated by the presence of Higgs and KK Higgs in the 
loops, rather than loops with KK gluons. The dependence with the KK
gluon mass scale is more pronounced for the tree-level decays, while in the loop
decays the decoupling with respect to the KK mass is less apparent, specially for
large Yukawa couplings and large Higgs localization parameter $a$. This can be
due to the parameter space points approaching a non-perturbative
regime, but it is  also partly due to the fact that the KK Higgs
masses happen to change less rapidly than the KK gluon masses as we
change the background parameters in order to increase the KK scale.

In summary, a comprehensive analysis of top flavor-changing neutral decays
in general warped extra dimensional models indicates that the only
decay with a chance to be observed is the tree-level decay  $t \to c
h$, while the loop-level decays  $t \to c\gamma$ and  $t \to c g$ seem
to fall well below the sensitivity of LHC @ 13 or 14 TeV.

\acknowledgments
MF  and MT thank NSERC and FRQNT for partial financial support under grants
number SAP105354  and PRCC-191578. ADF gratefully acknowledges the SNI-CONACyT and RXA acknowledges the CONACyT for the PhD fellowship.

\section{Appendix}
\label{sec:appendix}
We include here, for completeness the analytical expressions for the loop calculations presented in \ref{subsec:analytic}.

\subsubsection{$t\to q \gamma$, Higgs loop contributions, zero mode and KK}
\label{subsubsec:gamHiggs}

For $t_R$, charged Higgs
\begin{eqnarray}
F_{T_R}^{\gamma\,+}&=& \frac{e}{16 \pi^2}\sum \limits_{i=0}^{nKK} \sum \limits_{j=0}^{mKK} \frac{1}{M_{H^+_i}^2} \Bigg \{ \left [ m_t \left (Y^{H_i^+t_R D_j}\right) \left (Y^{H_i^+c_R D_j}\right)^{\star} + m_c \left (Y^{H_i^+t_L D_j}\right) \left (Y^{H_i^+c_L D_j}\right)^{\star} \right] \nonumber \\
&\times & \left [-\frac13 f_1(x_{ji})+g_1(x_{ji}) \right] + \left [M_{D_j} \left (Y^{H_i^+t_L D_j}\right) \left (Y^{H_i^+c_R D_j}\right)^{\star}\right ] 
\times 
 \left[ -\frac13 f_2(x_{ji})+g_2(x_{ji}) \right]  \Bigg \} \, .   \nonumber \\ 
\end{eqnarray}
For $t_L$, charged Higgs
\begin{eqnarray}
F_{T_L}^{\gamma\,+}&=& \frac{e}{16 \pi^2}\sum \limits_{i=0}^{nKK} \sum \limits_{j=0}^{mKK} \frac{1}{M_{H^+_i}^2}  \Bigg \{ \left [ m_c \left (Y^{H_i^+t_R D_j}\right) \left (Y^{H_i^+c_R D_j}\right)^{\star} + m_t \left (Y^{H_i^+t_L D_j}\right) \left (Y^{H_i^+c_L D_j}\right)^{\star} \right] \nonumber \\
&\times & \left[ -\frac13 f_1(x_{ji})+g_1(x_{ji}) \right] + \left [M_{D_j} \left (Y^{H_i^+t_R D_j}\right) \left (Y^{H_i^+c_L D_j}\right)^{\star} \right ] %
\times  \left[ -\frac13 f_2(x_{ji})+g_2(x_{ji}) \right]  \Bigg \}\, ,  \nonumber \\
\end{eqnarray}
where we used $\displaystyle x_{ji}= \frac{M^2_{D_j}}{M^2_{H^+_i}}$ and $nKK$ is the number of Higgs modes included, $mKK$ is the number of fermion modes included for each flavor.
\vskip0.5in

For $t_R$, neutral Higgs
\begin{eqnarray}
F_{T_R}^{\gamma\,0}&=& \frac{e}{16 \pi^2}\sum \limits_{i=0}^{nKK} \sum \limits_{j=0}^{mKK} \frac{1}{M_{H^0_i}^2} \Bigg \{ \left [ m_t \left (Y^{H_i^0 t_R U_j}\right) \left (Y^{H_i^0 c_R U_j}\right)^{\star} + m_c \left (Y^{H_i^0 t_L U_j}\right) \left (Y^{H_i^0 c_L U_j}\right)^{\star} \right] \nonumber \\
&\times & \left [\frac23 f_1(y_{ji}) \right] + \left [M_{U_j} \left (Y^{H_i^0t_L U_j}\right) \left (Y^{H_i^0 c_R U_j}\right)^{\star}\right ] 
\times  \left [\frac23 f_2(y_{ji}) \right] \Bigg \}  \, . \nonumber \\
\end{eqnarray}
For $t_L$, neutral Higgs
\begin{eqnarray}
F_{T_L}^{\gamma\,0}&=&\frac{e}{16 \pi^2} \sum \limits_{i=0}^{nKK} \sum \limits_{j=0}^{mKK} \frac{1}{M_{H^0_i}^2}  \Bigg \{ \left [ m_c \left (Y^{H_i^0t_R U_j}\right) \left (Y^{H_i^0c_R U_j}\right)^{\star} + m_t \left (Y^{H_i^0t_L U_j}\right) \left (Y^{H_i^0c_L U_j}\right)^{\star} \right] \nonumber \\
&\times & \left [ \frac23 f_1(y_{ji}) \right] + \left [M_{U_j} \left (Y^{H_i^0t_R U_j}\right) \left (Y^{H_i^0c_L U_j}\right)^{\star}\right ]
\times  \left [\frac23 f_2(y_{ji}) \right] \Bigg \} \, ,  \nonumber \\
\end{eqnarray}
where we used $\displaystyle y_{ji}= \frac{M^2_{U_j}}{M^2_{H^0_i}}$.
\vskip0.5in

\subsubsection{$t \to q \gamma$, KK Gluon Loop Contributions}
\label{subsubsec:gamgluon}

\begin{eqnarray}
F_{T_R}^{\gamma\,G}&=&C(R) \frac{e}{16 \pi^2} \sum \limits_{i=0}^{nKK} \sum \limits_{j=0}^{mKK}\frac{1}{M_{G_i}^2} \Bigg \{\left [ m_t  \left (\chi^{G_i U_j c_L}\right)^\dagger \left ( \chi^{G_i U_j t_L}\right) f_1(x_{ji})+ M_{U_j}  \left (\chi^{G_i U_j c_L}\right)^\dagger \left ( \chi^{G_i U_j t_R}\right) g_1(x_{ji}) \right] \Bigg \}\, ,  \nonumber \\
F_{T_L}^{\gamma\,G}&=&C(R) \frac{e}{16 \pi^2} \sum \limits_{i=0}^{nKK} \sum \limits_{j=0}^{mKK} \frac{1}{M_{G_i}^2} \Bigg \{ \left [ m_t  \left (\chi^{G_i U_j c_R}\right)^\dagger \left ( \chi^{G_i U_j t_R}\right) f_1(x_{ji})+ M_{U_j} \left (\chi^{G_i U_j c_R}\right)^\dagger \left ( \chi^{G_i U_j t_L}\right) g_1(x_{ji})\right] \Bigg \}\, ,  \nonumber \\
\end{eqnarray}
 where $\displaystyle x_{ji}= \frac{M^2_{U_j}}{M^2_{G_i}}$ and $\chi^ {G_i U_jc(t)_{L,R}}$ are the KK gluon couplings to $c(t)$ external quarks and KK $U$ fermions, and where and $C(R)=4/3$ is the quadratic Casimir operator on the fundamental representation of $SU(3)_c$.
We have
\begin{eqnarray}
C_{7\, \gamma}^L&=&F_{T_L}^{\gamma\,+}+F_{T_L}^{\gamma\,0}+ F_{T_L}^{\gamma\,G} \, , \nonumber\\
C_{7\, \gamma}^R&=&F_{T_R}^{\gamma\,+}+F_{T_R}^{\gamma\,0}+ F_{T_R}^{\gamma\,G}\, ,\nonumber
\end{eqnarray}
and 
\begin{equation}
\Gamma (t \to c \gamma)=\frac{m^3_t}{16 \pi} \left( |C_{7\, \gamma}^L|^2+|C_{7\, \gamma}^R|^2\right)\, .
\end{equation}


\subsubsection{$t\to q g$, Higgs loop contributions, zero mode and KK}
\label{subsubsec:gHiggs}

For $t_R$, charged Higgs
\begin{eqnarray}
F_{T_R}^{g\,+}&=& \frac{g_s}{16 \pi^2}\sum \limits_{i=0}^{nKK} \sum \limits_{j=0}^{mKK} \frac{1}{M_{H^+_i}^2} \Bigg \{ \left [ m_t \left (Y^{H_i^+t_R D_j}\right) \left (Y^{H_i^+c_R D_j}\right)^{\star} + m_c \left (Y^{H_i^+t_L D_j}\right) \left (Y^{H_i^+c_L D_j}\right)^{\star} \right] f_1(x_{ji})\nonumber \\
&+ &   \left [M_{D_j} \left (Y^{H_i^+t_L D_j}\right) \left (Y^{H_i^+c_R D_j}\right)^{\star}\right ] f_2(x_{ji})   \Bigg \}  \, .
\end{eqnarray}

For $t_L$, charged Higgs
\begin{eqnarray}
F_{T_L}^{g\,+}&=& \frac{g_s}{16 \pi^2}\sum \limits_{i=0}^{nKK} \sum \limits_{j=0}^{mKK} \frac{1}{M_{H^+_i}^2}  \Bigg \{ \left [ m_c \left (Y^{H_i^+t_R D_j}\right) \left (Y^{H_i^+c_R D_j}\right)^{\star} + m_t \left (Y^{H_i^+t_L D_j}\right) \left (Y^{H_i^+c_L D_j}\right)^{\star} \right] 
 f_1(x_{ji})\nonumber\\
 & + &\left [M_{D_j} \left (Y^{H_i^+t_R D_j}\right) \left (Y^{H_i^+c_L D_j}\right)^{\star} \right ] 
 f_2(x_{ji}) \Bigg \}   \, , 
\end{eqnarray}
where as before $\displaystyle x_{ji}= \frac{M^2_{D_j}}{M^2_{H^+_i}}$ and $nKK$ is the number of Higgs modes included, $mKK$ is the number of fermion modes included for each flavor.

For $t_R$, neutral Higgs
\begin{eqnarray}
F_{T_R}^{g\,0}&=& \frac{g_s}{16 \pi^2}\sum \limits_{i=0}^{nKK} \sum \limits_{j=0}^{mKK} \frac{1}{M_{H^0_i}^2} \Bigg \{ \left [ m_t \left (Y^{H_i^0 t_R U_j}\right) \left (Y^{H_i^0 c_R U_j}\right)^{\star} + m_c \left (Y^{H_i^0 t_L U_j}\right) \left (Y^{H_i^0 c_L U_j}\right)^{\star} \right] 
 f_1(y_{ji}) \nonumber\\
 &+&  \left [M_{U_j} \left (Y^{H_i^0t_L U_j}\right) \left (Y^{H_i^0 c_R U_j}\right)^{\star}\right ] 
 f_2(y_{ji}) \Bigg \} \, . 
\end{eqnarray}

For $t_L$, neutral Higgs
\begin{eqnarray}
F_{T_L}^{g\,0}&=&\frac{g_s}{16 \pi^2} \sum \limits_{i=0}^{nKK} \sum \limits_{j=0}^{mKK} \frac{1}{M_{H^0_i}^2}  \Bigg \{ \left [ m_c \left (Y^{H_i^0t_R U_j}\right) \left (Y^{H_i^0c_R U_j}\right)^{\star} + m_t \left (Y^{H_i^0t_L U_j}\right) \left (Y^{H_i^0c_L U_j}\right)^{\star} \right] 
 f_1(y_{ji})  \nonumber \\
 &+& \left [M_{U_j} \left (Y^{H_i^0t_R U_j}\right) \left (Y^{H_i^0c_L U_j}\right)^{\star}\right ]
 f_2(y_{ji})  \Bigg \}\, ,  
\end{eqnarray}
where as before $\displaystyle y_{ji}= \frac{M^2_{U_j}}{M^2_{H^0_i}}$.
\vskip0.5in

\subsubsection{$t\to q g$, KK Gluon Loop Contributions}
\label{subsubsec:ggluon}

\begin{eqnarray}
F_{T_R}^{g\,G}&=& \frac{g_s}{16 \pi^2} \sum \limits_{i=0}^{nKK} \sum \limits_{j=0}^{mKK}\frac{1}{M_{G_i}^2} \Bigg \{ m_t \left (\chi^{G_i U_j c_L}\right)^\dagger \left ( \chi^{G_i U_j t_L}\right) \left [-C(G) f_1(x_{ji}) + \left (2C(R)-C(G)\right ) f_2(x_{ji}) \right] \nonumber \\
&+& M_{U_j} \left (\chi^{G_i U_j c_L}\right)^\dagger \left (\chi^{G_i U_j t_R} \right)  \left [ -C(G) g_2(x_{ji}) + \left ( 2C(R)-C(G) \right ) g_1(x_{ji}) \right] \Bigg \} \, ,  \nonumber \\
F_{T_L}^{g\,G}&=& \frac{g_s}{16 \pi^2} \sum \limits_{i=0}^{nKK} \sum \limits_{j=0}^{mKK} \frac{1}{M_{G_i}^2} \Bigg \{  m_t \left (\chi^{G_i U_j c_R} \right)^\dagger \left ( \chi^{G_i U_j t_R}\right) \left [-C(G) f_1(x_{ji}) + \left (2C(R)-C(G)\right ) f_2(x_{ji}) \right] \nonumber \\
&+& M_{U_j} \left (\chi^{G_i U_j c_R}\right)^\dagger \left (\chi^{G_i U_j t_L} \right)  \left [ -C(G) g_2(x_{ji}) + \left ( 2C(R)-C(G) \right ) g_1(x_{ji}) \right] \Bigg \} \, ,
\end{eqnarray}
 where $\displaystyle x_{ji}= \frac{M^2_{U_j}}{M^2_{G_i}}$ and $\chi^ {G_i U_jc(t)_{L,R}}$ are the KK gluon couplings to $c(t)$ external quarks and KK $U$ fermions, and where and $ C(G)=3$ and $C(R)=4/3$ are the quadratic Casimir operator on the adjoint and fundamental representation of $SU(3)_c$, respectively.
\vskip0.5in

We have
\begin{eqnarray}
C_{8\, g}^L&=&F_{T_L}^{g\,+}+F_{T_L}^{g\,0} +F_{T_L}^{g\,G}\, , \nonumber\\
C_{8\, g}^R&=&F_{T_R}^{g\,+}+F_{T_R}^{g\,0} +F_{T_R}^{g\,G}\, ,\nonumber
\end{eqnarray}
and 
\begin{equation}
\Gamma (t \to c g)=\frac{m^3_t}{16 \pi} \left( |C_{8\, g}^L|^2+|C_{8\, g}^R|^2\right)\, .
\end{equation}

The loop functions in the above expressions are:
\begin{eqnarray}
f_1(x)&=&-\frac{x^3-6x^2+3x+2+6x \ln x}{12(x-1)^4} \, ,  \\
f_2(x)&=&-\frac{x^2-4x+3+2 \ln x}{2(x-1)^3} \, ,  \\
g_1(x)&=&\frac{2x^3+3x^2-6x+1-6x^2 \ln x}{12(x-1)^4} \, ,  \\
g_2(x)&=&\frac{x^2-1-2x \ln x}{2(x-1)^3} \, . 
\end{eqnarray}

\end{document}